\documentclass[12pt,a4paper]{article}
\usepackage{amssymb,amsmath}
\usepackage[dvips]{lscape,graphicx}

\newcommand{\bc}{\begin{center}}
\newcommand{\ec}{\end{center}}
\newcommand{\bd}{\begin{displaymath}}
\newcommand{\ed}{\end{displaymath}}
\newcommand{\be}{\begin{equation}}
\newcommand{\ee}{\end{equation}}
\newcommand{\ba}{\begin{array}}
\newcommand{\ea}{\end{array}}
\newcommand{\bt}{\begin{tabular}}
\newcommand{\et}{\end{tabular}}

\begin{document}


\title{Remarkable coincidence for the top Yukawa coupling and an approximately
massless bound state}

\author{C.~D.~Froggatt$^{a,b}$,
H.~B.~Nielsen$^{b}$ \\[5mm]
\itshape{$^a$ Department of Physics and Astronomy,}\\[2mm]
\itshape{Glasgow University, Glasgow G61 3JX, Scotland}\\[2mm]
\itshape{$^b$ The Niels Bohr Institute, Copenhagen 2100, Denmark}\\[2mm]
}

\date{}

\maketitle

\begin{abstract}{
\noindent
 We calculate, with several corrections, the nonrelativistic binding
by Higgs exchange and gluon exchange between six top and six
antitop quarks (actually replaced by left-handed $b$ quarks from
time to time). The {\em remarkable result} is that, within our
calculational accuracy of the order of 14\% in the top-quark
Yukawa coupling $g_t$, the experimental running top-quark Yukawa
coupling $ g_t = 0.935$ happens to have just that value which
gives a perfect cancellation of the unbound mass = 12 top-quark
masses by this binding energy. In other words the bound state is
massless to the accuracy of our calculation. Our calculation is in
disagreement with a similar calculation by Kuchiev et al., but
this deviation may be explained by a phase transition. We and
Kuchiev et al.~compute on different sides of this phase
transition.}
\end{abstract}

\newpage
\section{Introduction}
We have earlier claimed \cite{itepportoroz} that, if of the order
of 6 top quarks and 6 antitop quarks were brought within the
distance of the Higgs Compton wavelength from each other, they
would obtain such a strong binding energy that it would be
definitely of {\em the same order of magnitude as the mass energy
of these 6 top and 6 antitop quarks}. Within the accuracy of the
previous calculations \cite{itepportoroz,coral,pascos}, it was not
excluded that the binding energy for the 6 top plus 6 antitop
bound state could just compensate the mass energy, so that the
total mass of the bound state would be just zero. Indeed we
concluded that, within uncertainty, we had consistency with the
hypothesis that the experimental top-quark Yukawa coupling
constant $g_t$ had been ``mysteriously'' tuned so as to make this
bound state of 6 top and 6 antitop quarks---called NBS (new bound
state) or the t ball---have just zero mass. In our notation (see
Appendix A), the experimental top-quark Yukawa coupling is 0.935
corresponding to the running mass or 0.992 if we use the pole mass
of $172.6$ GeV \cite{mtop}.

However, it was recently claimed by Kuchiev, Flambaum and Shuryak
\cite{Shuryak} that, with the experimental coupling $g_t\approx
0.989$ and a realistic Higgs mass used to give the Yukawa form of
the potential, the system of the 6 top and 6 antitop quarks would
not even bind, let alone bind to zero mass. As we shall show in
Appendix J, the calculation of the bound state mass or rather the
lowest energy for the $6t + 6 \overline{t}$ system as a function
of the Yukawa coupling $g_t$ has the character of a ``phase
transition.'' That is to say that the mass $m_{bound}$ of the
bound state gets a kink as a function of the Yukawa coupling
$g_t$. In this Appendix we shall only present a toy model
calculation showing such a kink, but the expectation is that also
the fully correct calculation would at least approximately show a
kink. Having such a nonanalytic, or at least essentially
nonanalytic, behavior of the bound state mass in mind, it could
easily happen that the fact that Kuchiev et al.~ignore some
corrections could lead them to a qualitatively wrong conclusion by
working on the wrong side of the phase transitionlike kink. As we
illustrate with our toy model in Appendix J, the suggested phase
transition should be caused by the vacuum Higgs field collapsing
under the influence of the high density of top and antitop quarks
in the potential bound state.

Two calculations were made in Ref.~\cite{Shuryak} using,
respectively, a variational method and self-consistent Hartree
Fock equations. As we remark in Appendix I our own estimate of
what we call the many body effect agrees with the results of this
reference in the massless Higgs approximation. Kuchiev et
al.~effectively included $u$-channel Higgs exchange and also gluon
exchange, in addition to the explicitly considered $t$-channel
Higgs exchange, by increasing the $t$-channel Higgs exchange
potential by a factor of 2. However, in the present article we
want to take into account some effects not considered in Ref.~\cite{Shuryak}. 
The two  most important of these are as follows:

1) First, one has to take into account the possibility that, {\em
if the system of the 6 top plus 6 antitop quarks indeed binds
strongly}, then inside the bound state the Higgs field can be
strongly reduced compared to its usual vacuum expectation value
(VEV). Since the second derivative of the effective potential for
the Higgs field can even be negative, leading to an effectively
tachyonic Higgs particle, for small values of the Higgs field, it
follows that the effective Higgs mass inside the bound state might
be considerably smaller than the usual Higgs mass outside. This
correction of using an effective Higgs mass only sets in when
indeed one has the bound state. So it is {\it a priori} not
excluded that there could be a binding due partly to this effect,
while a calculation not using an effective Higgs mass might still
show no binding.

2) We shall take into account also the exchange of $W$ and $Z$ bosons
and even the photon. What really matter most in this connection are, as
we shall see below, the components of the intermediate gauge bosons
which in reality correspond to the eaten Higgs field components. So it is really
the exchange of components of the Higgs field, other than the radial
one identified with the genuine Higgs particles, which we consider here.

In addition we shall present several smaller corrections which
were not considered in Ref.~\cite{Shuryak}.

In the present article we correct and improve our previous
calculations \cite{itepportoroz,coral,pascos} of the ``critical''
top Yukawa coupling $g_{t}|_{phase \;transition}$ needed to make
the bound state, NBS or t ball or dodecaquark, massless. As
mentioned above we include first of all part of the effect on the
binding coming from the exchange of W and Z bosons, namely, the
part that is in reality the exchange of the ``eaten Higgses.''
Even in the limit of vanishing fine structure constants,
$\alpha_1$ for $U(1)$ and $\alpha_2$ for $SU(2)$, respectively,
there would be an exchange interaction between bottom and/or top
quarks by the exchange of W's or Z's. This is because the squared
masses in the propagators of the gauge bosons are proportional to
the fine structure constants and consequently, for part of the
exchange potential, the fine structure constants drop out of the
calculation. It is a part we interpret as being really the
exchange of an eaten Higgs component. Inclusion of at least part
of the W exchange means that a top quark gets converted to a $b$
quark or oppositely. Including such contributions, we thus have to
imagine that our NBS or t-ball bound state is a superposition not
only of top and antitop quarks, but also has components with some
of these top or antitop quarks replaced by $b$ quarks or anti-$b$
quarks, respectively. It is, however, trivial to see that the
right-handed chirality $b$ quark is totally decoupled in the first
approximation and cannot come into this approximation.

We consider the inclusion of this weak interaction exchange so to
speak---really of eaten Higgses---as a major correction, on top
of which we shall then further make a series of ``smaller''
corrections, which are typically really not so much smaller. The
whole calculation is basically a nonrelativistic one, although one
of our many corrections is an attempt to take into account in a
somewhat crude way relativistic effects.

The major purpose of these calculations is to see to what accuracy
we should be able to claim that the cancellation of the binding
energy and the energy of the masses of the constituents occur just
for the experimental top-quark Yukawa coupling. If we could claim
the accuracy is high enough, it would mean that nature had chosen
a very special value for the top-quark Yukawa coupling, and there
would be some mysterious fine-tuning problem to be explained---for
instance by our multiple point principle
\cite{itepportoroz,brioni}.

But even not worrying about the strange coincidence which such a
special top Yukawa coupling would mean, it would say that now the
bound state became of such low mass that there might be a hope of
producing it. In fact we expect a spectrum of bound states with
different numbers of top-quark constituents, as discussed in
Appendix H, and some of these should be found \cite{ichep08} at
the LHC and conceivably could already have been produced at the
Tevatron.

It should be emphasized that we {\em only use the standard model}
to calculate the critical Yukawa coupling needed to make the bound
state mass zero. If indeed the binding is small (or it does not
bind at all as Kuchiev et al.~\cite{Shuryak} would claim), even
after additional corrections the system will remain
nonrelativistic. However, if the binding gets of the same order as
the mass energy (12$m_t$), relativistic considerations are called
for. Kuchiev et al.~never need such relativistic corrections; but
we formulate our whole calculation the opposite way around, in as
far as we formulate it as calculating that specific value of the
top-quark Yukawa coupling $g_t$, which makes the bound state of
the 6 top and the 6 antitop quarks massless. Doing it this way
immediately forces us into the consideration of relativistic
calculations.

Our attempts to make the best relativistic estimate for instance
lead us to think of our calculation being done in the infinite
momentum frame (see Sec.~\ref{sec2} and Appendix B). In this frame
we find that the binding needed to make the bound state mass zero
is decreased by a factor of 2 compared to what a naive
nonrelativistic calculation would suggest. It corresponds to
extrapolating the mass squared linearly as a function of the
binding energy rather than extrapolating the mass. Also the
relativistic speeds mean that the exchanged Higgses cannot
necessarily be fully described by means of static Coulomb forces,
but that ladder diagrams with crossing rungs (lines) have to be
included (see Sec. \ref{finitespeed}).

In Sec.~\ref{sec2} we shall shortly review and update the
nonrelativistic calculation, formulated as a calculation of the
value of the top-quark Yukawa coupling $g_t$ needed for the
cancellation of the binding energy with the constituent masses. In
Sec.~3 we shall make the first crude introduction of the important
effect of including the eaten Higgs exchanges and the thereby
associated introduction of a component of left-handed $b$ quarks
in the bound state.

In the following sections we shall then go through several
``smaller'' corrections to our calculation: In Sec.~4 we make some
correction to our too crude treatment of the eaten Higgs exchange
force.

In Sec.~5 we discuss the correction of the Higgs mass after all
not being exactly zero, so that in principle we have a Yukawa
potential rather than, as we used at first, a massless Higgs
approximation meaning a Coulomb-like potential. However the point
will be that, due to the Higgs field from the many top quarks and
antitop quarks largely compensating or even overcompensating the
vacuum expectation value of the Higgs field, the effective
potential for the Higgs field has a second derivative
(corresponding to the effective mass squared) in the relevant
region which is even some times negative. It follows that an
effective Higgs mass, introduced to approximately describe the
relevant situation, will be much smaller than the physical Higgs
mass. For definiteness, in this paper we shall take the physical
Higgs mass to be $m_h= 115$ GeV, corresponding to the LEP lower
bound which coincides with the 2 standard deviation hint of a
Higgs signal \cite{mhlep}; this value is also consistent with our
multiple point principle prediction \cite{ichep08,fn2,fnt}.
However we shall also take a conservative error of $\pm 50$ GeV on
$m_h$.

In Sec.~6 we include the $s$-channel exchange of both the Higgs
and gluons; these contributions were left out in our previous
calculations, since they are more difficult to estimate and more
uncertain.

Then in Sec.~7 we consider the correction due to the dependence of
the top-quark mass on the value of the Higgs field inside the
bound state. A correction due to the finite speed of Higgs
exchange is considered in Sec.~8.

In Sec.~9 we correct for the very crude way in which we previously
treated the genuine many body problem, which occurs when we have a
system of 12 constituent particles. Previously we assumed that we
could calculate this effect by letting each top or anti-top quark
``feel'' the field of 11/2 of the other 11 particles, meaning that
there is on the average 11/2 other particles inside a sphere
around the center of mass point reaching out to the particle in
question and 11/2 outside. The field from the outside ones is
supposed to be negligible on the average, while that from the
inside ones can be treated as if they were all in a ``nucleus'' in
the center. As mentioned above this many body effect was also
calculated in Ref.~\cite{Shuryak}. Our results for the size of the
effect are in agreement with this paper for the massless Higgs
exchange approximation.

In Sec.~10 we consider the contribution to the binding energy from
the exchange of $SU(2)$ gauge bosons. Then in Sec.~11 we consider
$U(1)$-gauge boson exchange.

In Sec.~12 we discuss at what precise value of the scale $\mu$ we
think that our calculation delivers the top-quark running Yukawa
coupling $g_t(\mu)$. We call this the renormalization group
correction.

Then in Sec.~13 we present the final results and collect an
estimated uncertainty for the various corrections and thereby
essentially for the whole calculation.

In Sec.~14 we present the conclusion and discussion.

\section{Nonrelativistic binding of top and antitop quarks by Higgs
and gluon exchanges} \label{sec2} Calculating bound states in
relativity is generally difficult and in principle should be done
using the Bethe-Salpeter equation. However, it is much easier to
work with atomic physics, e.g., the Bohr atom, and the infinite
momentum frame technology, which to some extent has similar
simplifying properties to the nonrelativistic approximation, and
this is what we in principle shall use now. We shall use first the
nonrelativistic approximation and claim that, if we only consider
the $t$-channel exchange and only one constituent going around a
central particle or bunch of particles, we can simply use the old
Bohr formulas for the hydrogen atom if we ignore the mass of the
exchanged Higgs particle being different from zero---i.e., we use
a Coulomb potential rather than the true Yukawa one. But then we
argue that, for weak binding, we can trivially derive the binding
in the infinite momentum frame from the nonrelativistic one, so
that we can essentially use the Bohr formula also in the infinite
momentum frame. It is our intention next to include also
$u$-channel exchange while leaving the $s$-channel exchange, which
gets appreciably more complicated, to the later Sec.~\ref{s6}.

The virtual exchange of the Higgs particle between two quarks, two
antiquarks or a quark-antiquark pair yields an attractive force in
each case. For top quarks Higgs exchange provides a strong force,
since we know phenomenologically that $g_t(\mu) \sim 1$ in a
notation in which the Lagrangian density for the Higgs top-quark
interaction is $\frac{g_t}{\sqrt{2}}\overline{\psi}_{tL}
\phi_h\psi_{tR}+ h.c.$, where then the Higgs field is normalized
to the expectation value $<\phi_h>= 246$ GeV. See Appendix A for
our notation. In this notation the potential between two top or
antitop quarks, using only the $t$-channel exchange with massless
Higgs particles, is \be V_{t-channel \ Higgs} =
-\frac{g_t^2/2}{4\pi r}. \label{f1}\ee

So let us now consider putting more and more $t$ and
$\overline{t}$ quarks together in the lowest energy relative
s-wave states, the 1s wave. The Higgs exchange binding energy for
the whole system becomes proportional to the number of pairs of
constituents, rather than to the number of constituents. So {\em a
priori}, by combining sufficiently many constituents, the total
binding energy could exceed the constituent mass of the system.
However we can put a maximum of $6t + 6\overline{t}$ quarks into
the ground state 1s wave. We shall now estimate the binding energy
of such a 12 particle bound state.

As a first step we consider the binding energy $-E_1$ of one of
them to the remaining 11 constituents treated as just one
particle, analogous to the nucleus in the hydrogen atom but
consisting of $Z=11$ quarks.

However, if we want to be allowed to  sum the various $E_1$s
obtained for the 12 constituents, in order to obtain the total
potential energy of the system (as we must to calculate the bound
state mass), we must think of bringing the quarks or antiquarks
into the bound state one by one. That is to say that, when we
bring in the $i$'th constituent, the number of constituents in the
center is only $i-1$, so that the potential is
$-\frac{(i-1)g_t^2/2}{4\pi r}$. So, instead of taking the
potential felt by a single constituent in the final situation
(i.e., in the bound state)
 $V = -\frac{11g_t^2/2}{4\pi r}$, we take an average over the steps
 of putting in the particles one by one and use the potential:
\begin{equation}
V = -\frac{\frac{11}{2}g_t^2/2}{4\pi r}.
\label{Vt11}
\end{equation}

We assume that the radius of the system turns out to be reasonably
small, compared to the Compton wavelength of the Higgs particle,
and use the well-known Bohr formula for the ground state energy
level of a one-electron atom with atomic number $Z=11$, but
modified by the just mentioned inclusion of a factor $1/2$ {\em in
the potential} to obtain the crude estimate\footnote{This formula
actually represents a correction by a factor of 2 compared to our
previous publications \cite{itepportoroz,coral,pascos} in which we
instead divided the {\em binding energy} computed for $Z=11$ by 2;
but then one has forgotten that the factor $1/2$ in the potential
ends up being {\em squared} in the binding energy (the Rydberg in
Bohr's formula). This is  the effect of the average radius
increasing when the potential is decreased.}:
\begin{equation}
E_1 = -\left(\frac{\frac{11}{2}g_t^2/2}{4\pi}\right)^2 \frac{11m_t}{24}.
\label{binding}
\end{equation}
Here $g_t$ is the top-quark Yukawa coupling constant, in our
normalization in which the top-quark mass is given by $m_t = g_t
\, 174$ GeV. Furthermore we used the reduced mass of
$\frac{11}{12} m_t$ for the one top quark moving relative to the
other 11.

The nonrelativistic binding energy of the 12 particle system is
then given by  $E_{binding} = -12E_1$. This estimate only takes
account of the $t$-channel exchange of a Higgs particle between
the constituents.

\subsection{u channel}
\label{2p1} A simple estimate of the $u$-channel Higgs exchange
contribution \cite{itepportoroz} increases the binding energy by a
further factor of $(16/11)^2$. This can be seen as follows.
Considering that the system is totally antisymmetric in spin and
color permutations, we can effectively proceed as if it consisted
of 6 top quarks and 6 antitop quarks, with both of these bunches
being bosons. Then the permutation of the interacting particles
caused by going from $t$-channel to $u$-channel exchange means
adding them up as if the force were twice as big. Since the
considered quark can be permuted in this way with the remaining 5
quarks out of the other 11 quarks or antiquarks, we conclude that
the factor of 11 inside the square in Eq. (\ref{binding}) should
be replaced by $11+ 5 = 16$.  This gives:
\begin{equation}
 E_{binding} = -12 *(16/11)^2 E_1 =\frac{11g_t^4}{2\pi^2}m_t = 0.557 m_t g_t^4.
\label{binding2}
\end{equation}
Inclusion of the $u$-channel contribution in this way is equivalent
to using an averaged potential of
\begin{equation}
V_{with \ u-ch} = -\frac{\frac{16}{2} g_t^2/2}{4\pi r}.
\label{Vwithu}
\end{equation}

\subsection{Gluon exchange}
\label{gluon} We have so far neglected the attraction due to the
exchange of gauge particles. So let us estimate the main effect
coming from gluon exchange \cite{coral} due to the interaction
$g_s\overline{\psi_t} A^a_{\mu}\lambda^a/2 \gamma^{\mu}\psi_t$. It
follows that the $t$-channel gluon exchange graph gives an
effective Coulomb potential for a quark-antiquark pair in a color
singlet state of \be V_{gluon} = -\frac{g_s^2 Tr(\lambda^a/2*
\lambda^a/2)_{\underline{3}}} {4\pi Tr(I)_{\underline{3}}r} =
-\frac{g_s^2 8/2}{4\pi *3r} = -e_{t\overline{t}}^2/(4\pi
r)\label{Vgl}. \ee The QCD fine structure constant is given by
$\alpha_s(M_Z) = g_s^2(M_Z)/4\pi = 0.118$. However, as will be
discussed in Appendix C, the scale associated with the radius of
the new bound state is closer to the $m_t$ scale than to the $M_Z$
scale. We will therefore take the value $\alpha_S( m_t) = 0.109$
in our estimate. This corresponds to an effective gluon
$t-\overline{t}$ coupling constant squared of
\begin{equation}
e_{t\overline{t}}^2 = \frac{4}{3}g_s^2 = \frac{4}{3} 1.37 =  1.83.
\label{ett}
\end{equation}

Here, however, we must bear in mind that the gluon exchange
potential (\ref{Vgl}) is only for a quark attracting an antiquark
in the compensating color state. It is not the coupling between
all pairs of quarks and antiquarks; rather we should consider an
averaged gluon potential as follows.

For definiteness, consider a $t$ quark in the bound state; it
interacts with 6 $\overline{t}$ quarks and 5 $t$ quarks. The 6
$\overline{t}$ quarks form a color singlet  and so their combined
interaction with the considered $t$ quark vanishes. On the other
hand, the 5 $t$ quarks combine to form a color antitriplet, which
together interact like a $\overline{t}$ quark with the considered
$t$ quark. So the total gluon interaction of the considered $t$
quark is the same as it would have been with a single
$\overline{t}$ quark. In this case the $u$-channel gluon
contribution should equal that of the $t$ channel. We shall also
include a factor $1/2$, analogous to that included in the Higgs
potential $V$ above, which takes into account the probability of
the center of the effective 5 quark system being closer to the
center of the bound state than the considered quark. The averaged
gluon potential to be used, in analogy to the $t$ plus $u$ channel
Higgs exchange potential $ V_{with \ u-ch}$ of (\ref{Vwithu}),
thus becomes 2 times 1/2 of the expression (\ref{Vgl}), i.e.,
accidentally just $V_{gluon}$ itself. Thus the full averaged
potential, to be used as if all the quarks and antiquarks
interacted in the same way, is \be V_{total} = V_{gluon} + V_{with
\ u-ch} = -\frac{e_{t\overline{t}}^2}{4\pi r } -
\frac{\frac{16}{2} g_t^2/2}{4\pi r} = -\frac{e^2_{t \overline{t}}
+ 4 g_t^2}{4\pi r}. \label{Vtotal} \ee This means that the binding
energy (\ref{binding2}) should be corrected to include the gluon
exchange force by substituting \be 4g_t^2 \rightarrow  e^2_{t
\overline{t}} + 4 g_t^2, \ee which leads to (\ref{binding2}) being
replaced by
\begin{eqnarray}
 E_{binding} &=& \left(\frac{11(e^2_{t \overline{t}} + 4 g_t^2)^2}{32\pi^2}\right)m_t
\label{Ebinding} \\
 &=&  0.0348m_t (e^2_{t \overline{t}} + 4 g_t^2)^2\\
 &=& 0.557 m_t ( 0.456 +g_t^2)^2.
\label{bindingtot}
\end{eqnarray}
Later on, as we see that both the experimental $g_t$ value and the
critical $g_t$ value (which we are about to estimate) are close to
unity, it follows that as far as the coupling squared is concerned
the gluon potential is about half as strong as the Higgs
potential.

\subsection{Infinite momentum frame} \label{imf}
We can always think of our system as moving with a specified high
momentum in the $z$ direction. This is really considering the
infinite momentum frame. As long as the binding is so small that
higher order in it is irrelevant, we can trust the nonrelativistic
approximation and even translate it into infinite momentum frame
(IMF) language, in which the energy $E_{IMF}$  of a system of mass
$m$ having large momentum component $p_z$ and transverse momentum
$\vec{p}_T$ is generally written in the form \be E_{IMF} = p_z +
(m^2 + \vec{p}_{T}^2)/(2p_z). \label{IMF} \ee When we have an
object composed of several particles, each of them must have its
large momentum component $p_{zi} = x_i p_z$, where then $p_z$
stands for the total momentum of the cluster of particles in the
$z$ direction. In this notation the total infinite momentum frame
energy for such a cluster of $n$ constituent particles becomes \be
E_{IMF \ cluster} = p_z + \left (\sum_{i =1}^n (m_i^2 +
\vec{p}_{Ti}^2)/x_i \right )/(2p_z) + \mbox{interaction
terms}\label{EIMFc}. \ee

The reason we propose to think about this infinite momentum
frame---without even doing any proper calculation in it---is that
in this language we keep to the nonrelativistically looking
formula,\footnote{It is non-relativistic in the sense that the
kinetic term is quadratic in $\vec{p}_{T}$ \cite{susskind}.} as
long as $p_z$ is very large, even if the mass squared
$m_{bound}^2$ of the bound state\footnote{The mass squared
$m_{bound}^2$ of the bound state is defined such that, if the
constituent wave function corresponding to the bound state NBS is
used for the cluster (\ref{EIMFc}) and we put $E_{IMF} = E_{IMF \
cluster}$ into (\ref{IMF}), we get $m_{bound}^2 = m^2$.} we wish
to study should become small. Then we can, namely, imagine that
one can calculate the energy $E_{IMF \ bound}$ of the bound state,
due to the Higgs and gluon exchange, in the nonrelativistic way
even when the mass squared $m_{bound}^2$ is close to zero.
Supposing that this can be  done in a formalism in which one has a
Hamiltonian involving the $x_i$'s and their conjugates, as well as
the transverse momenta and their conjugates, we should expect the
$E_{IMF \ bound}$-energy eigenvalue to be analytic as a function
of the parameters. Hence $m_{bound}^2$, which is linearly related
to this energy, should also be analytic. Ignoring in such a
calculation higher order terms in the coupling $g_t$ than the
fourth order, which  we just used, we can then reliably get the
mass squared of the bound state $m_{bound}^2$ even become
negative, provided that the interaction is sufficiently strong.

That is to say that we can now obtain a tachyonic bound state with
$m_{bound}^2 < 0$. In this way  a new vacuum phase could appear
due to Bose-Einstein condensation. Let us consider a Taylor
expansion in $g_t^2$ for the mass {\em squared} of the bound
state, estimated from our nonrelativistic binding energy formula:
\begin{eqnarray}
m_{bound}^2 & = & \left(12m_t\right)^2 - 2\left(12
m_t\right)\times
E_{binding} + ...\label{e10}\\
& = & \left(12m_t\right)^2\left(1 -\frac{2*0.557 (0.456
+g_t^2)^2}{12}+
...\right) \\
& = & \left(12m_t\right)^2\left(1 -0.0929*(0.456 +g_t^2)^2 +
...\right ). \label{expansion}
\end{eqnarray}
Assuming that this expansion can, to first approximation, be
trusted---as our argument using the infinite momentum frame was
meant to suggest---for large $g_t$, the condition $m_{bound}^2=0$
for the appearance of the above phase transition with degenerate
vacua becomes to leading order\footnote{Due to the already
mentioned mistake in previous publications
\cite{itepportoroz,coral,pascos} by a factor 2 in the binding
energy, the incorrect value $g_t|_{phase \ transition}\simeq 1.24$
was previously quoted.}: \be
 0= 1 -0.0929*(0.456 +g_t^2)^2 \label{f17}
\ee
or
\begin{equation}
\label{gtphase} g_t|_{phase \ transition}\simeq
\sqrt{\sqrt{1/0.0929} - 0.456} = 1.68.
\end{equation}

At this level we have included $t$- and $u$-channel Higgs and
gluon exchange, both taken as massless particles, and we have only
used as constituents the top and antitop quarks. We have not
included the W or eaten Higgs exchange that could lead, as we
shall see below, to partly $b$ quarks among the constituents. Also
we worked in the nonrelativistic or infinite momentum frame
approximation, and so far we did not specify at what scale to take
the $g_t$ although we have put in essentially the perturbative QCD
scale at $m_t$ for the gluon coupling.

In the article of Kuchiev, Flambaum and Shuryak \cite{Shuryak},
these authors crudely take into account gluon exchange and
$u$-channel exchange by taking the potential between two top
quarks to be twice as big as the $t$-channel Higgs potential
(\ref{f1}). Using this we can extract from Eq.~(5) in their
article the critical value of $g_t$ needed to make the binding
energy $<-H>$ equal to just half of the mass energy $Nm_t$ (where
$N=12$ is the number of constituents)---as is required according
to our infinite momentum frame formula (\ref{factorhalf}) below.
This critical value becomes  $g_t|_{KFS}=1.91$. This value should
be compared to the just given value of 1.68 in (\ref{gtphase})
after it has been corrected for the many body effect performed in
Appendix I, which gives $1.68*(2.16)^{1/4} = 2.04$ to be compared
with $g_t|_{KFS}=1.91$. The small 6\% difference is mainly due to
the crude treatment of the gluon and $u$-channel correction.
Rather than an increase in the $t$-channel Higgs potential by a
factor of 2, we find that with the value (19) above for $g_t$ the
factor would be 1.70. With such a correction factor multiplying
the $t$-channel Higgs potential, we get $g_t|_{KFS}=
1.91/\sqrt{1.70/2} = 2.07$ in close agreement with our many body
corrected value of 2.04.

\subsection{Justification of formal nonrelativistic mass-energy cancellation
of half constituent mass to get zero mass bound state} \label{justification}

We already argued that analyticity of the energy in the infinite
momentum frame, or equivalently the bound state mass squared,
suggested that we could formally use the nonrelativistic binding
energy calculation and extrapolate it until the mass squared of
the bound state becomes zero or even less than zero, if we wanted
to obtain the phase transition value $g_t|_{phase \ transition} $
of the Yukawa coupling.

It is easy to see that the formal nonrelativistic requirement for
this extrapolation in mass squared to make $m_{bound}^2$ zero
means that the binding energy is adjusted to obey \be
 \sum_i m_i/2 - E_{binding} =0, \label{factorhalf}
\ee rather than the intuitively expected requirement, which does
not have the factor 1/2 on the mass term. This factor 1/2 came in
from the Taylor expansion (\ref{e10}) of $m_{bound}^2$ in terms of
the binding energy $E_{binding}$.

We would now like to justify such a formal rule for calculating
the critical $g_t$-value $g_t|_{phase \ transition} $. For that
purpose, in Appendix B, we imagine writing the infinite momentum
frame energy $E_{IMF \ cluster} $ for a cluster of
``constituents'' first in the case that the nonrelativistic
approximation is valid. In this case we obtain the energy
expression\footnote{Note that this expression (\ref{eq25}) agrees
with Eq.~(\ref{eq128}) of Appendix B, when all the constituents
move with the same speed in the $z$-direction so that $x_i =
\frac{m_i}{\sum_j m_j}$.} \be E_{IMF \ cluster } = p_z +
\frac{1}{2p_z} \sum_i \frac{m_i}{x_i}( m_i + 2H_i). \label{eq25}
\ee Here $H_i$ is the contribution of the $i$'th particle to the
total nonrelativistic Hamiltonian $H = \sum_i H_i$ in the rest
frame of the bound state:
\begin{equation}
  H_i = \frac{\vec{p}_i^2}{2m_t} + \frac{1}{2} \sum_{j, \ i\neq j}V_{ij}
\approx \frac{\vec{p}_i^2}{2m_t} + V_{total},
\label{Hi}
\end{equation}
where we approximate the interaction by a central potential
$V_{total}$, while only half of the other 11 particles are
present, and also include the $u$-channel and gluon exchange
contributions by using Eq.~(\ref{Vtotal}). This approximation
corresponds to taking the effective two particle interaction to be
\be V_{ij} \approx -\frac{A}{4\pi r_{ij}} \qquad \mbox{where}
\quad A = \frac{2( e^2_{t \overline{t}} + 4 g_t^2)}{11}.
\label{Vijtotal} \ee Here $r_{ij}$ denotes the distance between
the $i$th particle and the $j$th particle.

We could now imagine that we want to use the expression
(\ref{eq25}), in order to obtain the critical value for the Yukawa
coupling from the requirement of the bound state being of zero
mass. This would mean that the term proportional to
$\frac{1}{2p_z}$ should be zero to determine $g_t|_{phase \
transition}$. A symmetry argument between the different
constituents---at least in the case of interest here in which all
the constituents are the same type of particle---would suggest
that we have to obtain this zero by all the operator factors $ m_i
+ 2 H_i $ being actually zero, in the sense of the nonrelativistic
single particle Hamiltonians $H_i$ having eigenvalues $-m_i/2$. If
we believe this basic analyticity assumption argument, we have
arrived at a justification for our rule of calculating the
critical $g_t$ coupling, according to which one shall require
there be an eigenvalue for the binding energy equal to {\em half
the mass $m_i$ value}. In other words one shall find an
``eigenstate'' $\Psi$ for which the equations \be ( \frac{m_i}{2}
+ H_i) \Psi =0  \; \mbox{ for} \  i = 1, 2, ..., n \label{e18} \ee
are satisfied. Let us here stress that, in this equation
(\ref{e18}), the central potential approximation which we use in
the single particle Hamiltonian $H_i$ corresponds to the
half-filled situation, so that $Z=\frac{11}{2}$ and the potential
$\frac{1}{2}\sum_{j \ j\neq i} V_{ij}$ is replaced by $V_{total}$.

If one wanted the physical Higgs field, one should rather ask for
the Higgs potential felt by a constituent after all the other 11
constituents have already been put into the system. This would, in
our concentration in the center approximation, give twice as
strong a Higgs field as if one naively used the Higgs potential in
$H_i$. But this is only true for the large $r$ region, where one
truly can expect all the particles generating the Higgs field to
be at smaller $r$ than the considered one. So for large $r$ indeed
one should multiply the deviation of the Higgs field from the
usual VEV, as given by the Higgs potential in $H_i$, by a factor
of 2. However, for an average distance $<r>$, about half the field
producing constituents are farther away from the center and their
contribution can crudely be ignored. So for $r \approx <r>$ this
factor of 2 is compensated by the factor of 1/2 corresponding to
only getting the field from the constituents closer to the center.
So here you get the Higgs field corresponding to the Higgs
potential as present in the expression $H_i$. Further inside,
corresponding to $r$ less than $<r>$, the true field deviation
from the usual VEV is even smaller than that corresponding to the
Higgs potential in $H_i$.

The outcome of this discussion is that:
\\
1) We argue that it is reasonable to use the nonrelativistic
approximation as a rule that should lead to our wanted
calculation.
\\
2) It is important that in this rule one should get the zero mass
bound state by requiring only half the mass be compensated by the
binding. The other half should then in reality be canceled by the
suggested analytic extrapolation.

\section{Introduction of left-handed bottom quarks}
\label{sec3} We have so far left out the exchange of the weak
gauge bosons but, with the estimate of the radius of the bound
state given in Appendix C being of the order of $r_0 \simeq
(\sqrt{4/3}m_t)^{-1}$, we should not necessarily ignore weak gauge
boson exchanges. Actually in this section we shall
only include these weak gauge boson exchanges in the approximation
of letting the gauge couplings go to zero. At first you might
think that in this limit we could totally ignore the exchange of
the $W$ and $Z^0$, but that is not true, because for the
longitudinal components of these bosons there is then a zero in
the inverse propagator due to gauge symmetry. In fact these
longitudinal $W$ and $Z^0$ components really represent the
``eaten'' Higgs components (see Appendix D). So what we shall
really do is to replace the exchange of $Z^0$ and $W$ by the
exchange of their longitudinal components and postpone the
discussion of the effect of their Coulomb fields until Secs.~\ref{s10} 
and \ref{s11}. Equivalently we can think of it as the
exchange of the components of the Higgs other than the physical
Higgs particle, which we already considered at length in the
foregoing section.

In as far as the Higgs field has two complex components, of which
we have in the foregoing section only included the uneaten real
part of one of them, there are three more real fields in the Higgs
doublet, and these can be exchanged between the constituents in
our bound states. The components eaten by the $W$ are the charged
fields and they will when exchanged from a top quark convert it to
a $b$ quark or oppositely.

For the understanding of the correction due to these exchanges,
let us first note that we should have in mind that the particles
which couple sufficiently strongly to be included in our
approximation are as follows:
\\
1) The {\em left-handed} $b$ quark components.
\\
2) The whole Higgs doublet (but we do not need the $W$ and $Z^0$
fields proper, nor the photon field). Only the eaten components
and the physical Higgs are included.
\\
In this way we cannot properly have $b$ quarks or anti-$b$ quarks
in our bound states, since they would all the time have to be
represented by the right-handed top components whenever they need
to be represented by right-handed components.

We shall further make the approximation that, whenever a pair of
(left) $ b \overline{b}$ quarks has been made by eaten charged
Higgs exchange, then it soon gets again annihilated back to the
usual situation of there being only top quarks and antitop quarks
in our bound state. This assumption means that we only take into
account that a right-handed top and an anti-right-handed
top---which really must have left helicity---exchange an eaten
Higgs between them and become a $b \overline{b}$ pair of bottom
quarks, which then in the next interaction return back to become
again a pair of right-handed top quarks.

For definiteness one could think of the self-energy diagrams for a
combination of fields with the appropriate conserved quantum
numbers to have an overlap with the bound state. Then, due to the
summation over an infinite number of diagrams, a pole should be
generated at $p^2 = m_{bound}^2$ corresponding to the mass squared
of the bound state. In this section we shall use a box-diagram
approximation, according to which the dominant self-energy
diagrams are the ones in which the doublet propagators are
restricted to circulate around box subdiagrams, with singlet
right-handed top-quark propagators attached to the four vertices.
The singlet right-handed top-quark propagators are not restricted
and could, for instance, cross over each other forming a nonplanar
diagram.

In the previous section we only included the ``physical'' Higgs
component and only the left-handed top quark as particles that
could come into these box diagrams. So we could think of the
previous calculation as having used, for the box-diagram
description of the scattering of right-handed top quarks, only
those box diagrams in which the left-handed top quark and the
physical Higgs components were included. Since the physical Higgs
component is only one purely real part of one of the complex
components of a Higgs doublet (which has two complex components
equivalent to four real ones---meaning two purely real and two
purely imaginary components), we must also imagine that in a
corresponding sense in this approximation of the previous section
only the real part of the left-handed top-quark field was used.

By the inclusion of the left-handed bottom quark components and
the three eaten real components of the Higgs, we have 4 times as
many real components to exchange and to be represented as
propagators in the box diagram we described. Actually all four
combinations of Higgs and of left-handed bottom or top-quark real
components that can circle around in the box diagram will give the
same diagram contributions. Thus the effect of including the three
eaten components and the left-handed quark components connected
with them in the box diagram should simply be to increase the size
of the box-diagram contribution by a factor 4, as may be checked
in Appendix E. Now this box diagram is proportional to $g_t^4$,
because each of its four corners contributes a top-quark Yukawa
coupling. In Sec.~\ref{imf} we estimated the top-quark Yukawa
coupling needed to make the $6t + 6\overline{t}$ bound state to
have zero mass to be $g_t = 1.68$. Naively, in our present
approximation, this estimation could in principle have been made
by using only the box diagram, but really to claim that would need
some argumentation which we postpone to Sec.~\ref{s4}. This means
that, in this approximation, the contribution to the binding
energy from the box diagram with physical Higgs exchange would
have to have the same value as in the previous section, in order
to make the bound state massless. However we have seen that this
box-diagram contribution should be increased by a factor of 4.
This must mean that we should correct our predicted value of
$g_t^4$ down by a factor 4, in order to obtain again, in the more
correct calculation including the eaten Higgses, the massless
bound state of $6t + 6\overline{t}$.

Thus we have now reached the estimate that the critical coupling
$g_t$ arranged to make the proposed bound state of $6t +
6\overline{t}$ to have zero mass becomes \be g_t|_{phase \
transition} = 1.68 / 4^{1/4} =1.19. \label{bcorrected} \ee This
estimate of the Yukawa coupling, giving the exact masslessness of
the bound state of $6t + 6\overline{t}$, was made using only the
$t$- and $u$-channel gluon and Higgs exchanges. However, we made
an oversimplified approximation with respect to the exchange of
the eaten Higgses (really longitudinal $W$ and $Z^0$ exchange),
meaning that we considered deviations from there being only
physical Higgs exchange inside certain box diagrams. A major point
is that we have included the presence of left-handed $b$ quark
components as constituents rather than only top quarks. So we
should perhaps not say that our state is exactly composed from $6t
+ 6\overline{t}$, since actually it is now considered possible to
virtually replace a top and anti-top-quark pair by a bottom and
antibottom quark pair.

It is now the idea to make a series of smaller corrections below
to the approximations used to reach Eq.~(\ref{bcorrected}).

First, in Sec.~\ref{s4}, we shall discuss the correction to our
box-diagram approximation coming from other closed loops of weak
isospin doublet lines. However, before doing so, we consider the
possible effect of diagrams involving interactions with the VEV of
the Higgs field, represented by a Higgs-propagator symbol with a
cross at one end (a tadpole). Because weak isospin is formally
upheld in the Feynman rules, it follows that the couplings with
the Higgs VEV have to come in pairs. We here want to argue that,
when we precisely require the bound state to be exactly massless,
these diagrams involving tadpoles must add up to zero.

An argument for this runs as follows: Clearly the sum over those
diagrams having just two vacuum couplings will be proportional to
the square of the Higgs VEV. Provided we ignore the direct
dependence of the mass of the bound state on the Higgs mass (which
according to Sec.~5 contributes a $5.2 \%$ correction to
$g_t|_{phase \ transition}$), we expect, for dimensional reasons,
that the bound state mass for fixed values of the coupling
constants must be proportional to the Higgs VEV, except perhaps
for very small renormalization group effects. But now, as we
insist on looking for the zero mass case, there will be no
dependence on the Higgs VEV. In turn that means that the total
contribution to the change in the mass of the bound state, arising
from the diagrams with two tadpoles, must be just zero.

Accepting this argumentation then the contributions arising from
the insertion of one pair of tadpoles into the diagrams should at
the end add up to just zero. Really you can argue similarly that
the diagrams with four tadpoles and so on would also cancel out.

Finally we have argued that, for our specific project of finding
that $g_t$ value for which we can have a massless bound state, we
can ignore the tadpole diagrams and thus concentrate on those
diagrams in which all the isospin doublet propagators form loops
of longer or shorter lengths. We assumed above that it is the very
shortest loops which matter most, but in the next section we shall
discuss corrections to this box-diagram approximation.

\section{Corrections to the eaten Higgs exchange force and thereby bottom
quark admixture} \label{s4} Actually it is not correct that the
left-handed bottom or top quarks would only circle around in box
diagrams. In order to obtain an idea as to how much this
box-diagram approximation has to be corrected, let us imagine a
diagram being written down for how the bunch of 12 top or antitop
quarks propagate with mutual interaction under what really
corresponds to the development of the bound state. As in Sec.~2,
there are interactions between any of the top or antitop
constituents and any other one among them. We imagine constructing
the diagram by drawing a series of 12 top-quark lines representing
chains of top-quark propagators. Next we divide these lines up
into propagators while decorating them with exchanged particle
propagators going from one of the lines in the chains to one of
the other ones. At first we imagined that we had top-quark
propagators representing both right-handed and left-handed
components. But it is actually rather easy to imagine that, in our
Feynman rules, we make different propagators for right-handed and
left-handed components so as to introduce one propagator for left
and another one for right. Then the Higgs vertices all the time
connect a left to a right, while the gluon vertices oppositely
couple left to left and right to right.

After having imagined the notation with left-handed and
right-handed $t$ propagators being treated as different particles,
we can rather easily introduce the left-handed $b$ quarks by
imagining that we allow the left propagator to be treated as if it
had both a $t$ and a $b$ component built into it. So the left
propagators represent simultaneously two types of particles, $b$
and $t$, while the right propagators are kept unchanged and only
represent the right-handed $t$ quark. At the same time we have to
introduce also the eaten components for the Higgs propagator, but
that we do analogously by just deciding that now the
Higgs-propagator symbol stands for both complex components being
propagated. In other words we just reinterpret the diagram to
include the eaten components and the left-handed $b$ quark also.
The diagram will look formally the same as when we just separated
the diagram into left and right $t$ propagators without any $b$
quarks.

In this latter notation we can follow the propagation of the
doublets through the diagrams. That is to say we can follow chains
of propagators, which are either left $t$ and $b$ combined
propagators or the full Higgs propagators. Since these two types
of propagators are doublets under weak isospin, while the right
$t$ propagators and the gluons are weak isospin singlets, it is
clear that the chains of doublet propagators cannot end anywhere
in the interior of the diagram. Ignoring the case of external
lines being doublets, they would have to form loops inside the
diagram considered. In Sec.~3 we actually made the approximation
that these loops of doublets would always be box loops having only
four propagators along the loop. But that is of course by no means
guaranteed.

There is however an argument that the small box loops of doublets
might be favored compared to more extended doublet loops: We get
our factor 4 increase in the value of the whole diagram, due to
the inclusion of the eaten components and the $b$ quark, for each
doublet loop that can be found in such a whole diagram. For a
given number of left and Higgs propagators one thus gets the
biggest increase factor---i.e., more factors of 4---by putting the
doublet propagators into as many loops as possible. That will then
mean to put them into loops with as few propagators as possible
around them. But such loops correspond to box loops. Since the
propagators around a doublet loop must, namely, alternate a
left-handed quark with a Higgs back and forth, we must always have
an even number of propagators in such a loop of doublets. So four
is the minimum nontrivial number of propagators in the loop.

Were it not for such an effect of a somewhat higher factor for the
small doublet loops, the doublet loops could be rather long because
they would be obtained by combinatorically taking random diagrams.

We now want to correct for the fact that our assumption of there
being only the box loops of doublets overestimates the effect on
the correction to $g_t|_{phase \ transition}$ from the inclusion
of $b$ quarks. In fact we used above that the factor 4 per doublet
loop could be compensated for by a corresponding reduction in
$g_t^4$ by the factor 4. But now, since many of the doublet loops
can be longer than 4 propagators around but rather on the average
$n$ propagators around, the correction should instead have been
that $g_t^n$ be reduced by a factor 4. That of course would lead
to the change \be g_t|_{phase \ transition} = 1.68/
4^{1/n},\label{e28} \ee where we now have to estimate an
appropriate average for the quantity $n$.

A doublet loop with $n$ vertices along it has $n$ doublet
propagators. So, compared to the box doublet loops, it has per
loop $(n-4)/4 $ too few factors of 4 due to the summing over the
different components that can propagate around the loop. This
gives for such doublet loops a suppression weight factor
$4^{-(n-4)/4}$. This means that if you compare the contribution
for one diagram and one modified locally in the diagram
reorganizing it so as to replace $n/4$ box loops by one
$n$-``propagator'' loop\footnote{Here we count a series of doublet
propagators, which reduce to a single propagator when gluon
propagators are ignored, as a single ``propagator''. Really the
easiest way of thinking about this is to say that we totally
ignore the gluons in this calculation of the backcorrection to our
box-diagram approximation of Sec.~\ref{sec3}.} of isodoublets in
the local region considered, then the magnitude of the square of
this modified diagram will be $(4^{-(n-4)/4})^2$ times the
corresponding square of the replaced diagram. Let us suppose that
statistically, ignoring the extra factors for the isodoublet
loops, the distribution in a random (typical) diagram of the loop
size $n$ is smooth. This distribution of the number of propagators
around the isodoublet loops is briefly discussed in Appendix F.
Taking this distribution to be flat and essentially constant for
the first few $n$ values, we obtain that the probability
distribution of $n$ (on random diagrams weighted with their
magnitude squared) would go as $(4^{-(n-4)/4})^2$. If you somehow
weighted with amplitude rather than the squared amplitude, the
``distribution'' would only go as $4^{-(n-4)/4}$.

We may see that this means, in the example of, e.g., the six sided
doublet loop, that its weight factor is $4^{-(6-4)/4}$ = $1/2$ in
amplitude. But in probability the six sided loops are suppressed
rather by $(4^{-(6-4)/4})^2$ = $(1/2)^2$ = $1/4$. Thinking of the
Feynman diagrams as adding up with {\em random phases}, the
resulting sum of a lot of Feynman diagrams would statistically get
a magnitude corresponding to adding them in quadrature rather than
simply adding real positive numbers with the size of the series of
diagrams. We indeed take it that the weighting of the importance
of loops of a given number of doublet propagators $n$ shall be
counted as proportional to the squared quantity $(4^{-(n-4)/4})^2$
rather than to $4^{-(n-4)/4}$ itself. It is then easily seen that
the relative importance of the contributions of the loops with the
series of $n$ values (being $n= 4, 6, 8, 10, ...$) form the series
of terms \be 1 + \frac{1}{4} + \frac{1}{16} + \frac{1}{64} +
\cdots = \frac{4}{3}. \label{series} \ee It follows that, instead
of all the correction factors to $g_t|_{phase \ transition}$ in
Sec.~\ref{sec3} being $4^{-1/n} = 4^{-1/4}$ (coming by thinking of
just box loops having $n=4$), we get the following series of
correction factors corresponding to the series of terms in
(\ref{series}) \be 4^{-1/4} \ , \ 4^{-1/6} \ . \ 4^{-1/8} \ , \
4^{-1/10} \ , \ \cdots. \ee Compared to the correction as made in
Sec.~\ref{sec3} to $g_t|_{phase \ transition}$ (multiplicatively),
we get instead a further correction---which is really correcting
back for the fact that we have overcorrected---by \be \frac{1 * 1
+ \frac{1}{4} * 4^{-1/6+1/4} + \frac{1}{16} * 4^{-1/8 + 1/4} +
\cdots} {1 + \frac{1}{4} + \frac{1}{16} + \frac{1}{64} + \cdots }
\approx 1.04. \label{corf} \ee

We have thus crudely estimated that the $b$-inclusion correction
of Sec.~\ref{sec3} has to be modified, so as to put the estimate
of the critical $g_t|_{phase \ transition}$ {\em up} by 4\%.

For a very big number $N$ of Higgs components to
exchange and a corresponding number of left-handed quark
components, the smallest closed loop for the circulation of the
weak isospin will be favored. This is because we get a factor of
$N$ for each closed loop. Then for a self-energy diagram with a
given number of doublet propagators, we get the largest number of
factors of $N$ by using the diagram with the largest number of
loops. For large $N$ it follows that the box-diagram approximation
will dominate. This number $N$ is 4 in the true standard model as
already discussed above. For $N$ being small, however, we have no
guarantee for this box-diagram dominance at all and indeed this is
the situation for which our calculations in Sec.~\ref{sec2} were
performed, namely, for $N=1$. So we do not yet really have a good
argument for how this Sec.~\ref{sec2} calculation can be related
to the higher $N$ cases.

Let us now consider the dependence of $g_t|_{phase \ transition}$
on N, by introducing a parameter $n(N)$ giving the typical doublet
loop size in our complicated Feynman diagrams, so that \be
g_t|_{phase \ transition} = g_0/ N^{1/n}.\label{e29} \ee Here
$g_0$ does not depend on $N$. For large $N$ the box diagram
dominates and we clearly have $n(N) \rightarrow 4$. It is also
clear that the denominator  $N^{1/n} \rightarrow 1$ for $N=1$. But
then we can essentially {\em interpolate} the denominator to be
approximately $N^{1/4}$ for all $N$. Therefore we can effectively
calculate {\em as if} Eq.~(\ref{e29}) were replaced by \be
g_t|_{phase \ transition} \approx g_0/ N^{1/4}.\label{e294} \ee
and thereby justify Eq.~(\ref{bcorrected}) as a good
approximation, because it is a good interpolation of the general
formula.

We estimate that the uncertainty in this interpolation formula is
of order $\pm 7\%$. Combining this $\pm 7 \%$ error with an
estimated error of $\pm 4\%$ on the correction calculated above,
we obtain a total error of $\pm 8 \%$ Thus our final result for
the correction to $g_t|_{phase \ transition}$ is to increase it by
$4 \pm 8 \%$.

\section{Correction due to Higgs mass} \label{sec5}

In Appendix C we estimate the Bohr radius of our bound state of
$6t + 6\overline{t}$ in the critical coupling case to be $r_0
\approx (\sqrt{4/3} m_t)^{-1}$ and thus we see that with a Higgs
mass of 115 GeV, which we use in this paper, the effect of the
Higgs particle having a nonzero mass would not be so dramatic for
our calculations. It is however not just this argument of the
radius being small which is the true reason for our correction,
due to the nonzero  mass of the Higgs, being only a small
correction. Rather the argumentation is the following:

As we go into the interior of the bound state we find the Higgs
field due to all the top and antitop quarks around. These fields
have such a sign as to mean that, in reality, the normal vacuum
value of the Higgs field is  diminished in the interior of the
bound state. So we shall not use the Higgs mass for the normal
vacuum, but rather some effective Higgs mass on the background of
the Higgs field inside the bound state. This effective Higgs mass
squared is extracted from the second derivative of the effective
potential for the Higgs field at the value at which we want to
``work.'' To a good approximation the Higgs field effective
potential is given as a fourth order polynomial \be
V_{eff}(\phi_h) =- \frac{1}{2} |m_{h bare}|^2|\phi_h|^2  +
\frac{\lambda}{8}|\phi_h|^4. \ee So the physical Higgs mass
squared is related to the second derivative of this expression at
the minimum, where the value of the field $\phi_h$  must be fitted
to  $246$ Gev.

But now it is obvious that the second derivative and thus the
effective Higgs mass squared becomes smaller for lower values of
the field $\phi_h$, where we want to extract this second
derivative. Since in the interior of the bound state the Higgs
field is supposed to be smaller, then the Higgs mass to be used
there actually also becomes smaller than in the normal vacuum
outside the bound state. In this section we shall at first ignore gluonic
contributions to the binding energy. Then we estimate that the
Higgs field inside the bound state deviates so strongly from the
one in the normal vacuum that even the sign of the full Higgs
field---the vacuum value plus the field contributed by the
quarks---tends to be inverted in the most interior part of the
bound state. Thus, in most of the interior of the bound state
volume, the second derivative is much smaller than in the normal
vacuum or even negative. The latter corresponds formally to an
imaginary effective Higgs mass. So in an averaged way the Higgs
mass squared is, to first approximation, an average of both
negative and positive contributions.

Really we should split up the volume of the bound state into a
region---the more interior region---with imaginary effective Higgs
mass and a more exterior region in which the Higgs mass is real,
but even there the numerical value of the mass is diminished.

In order to get an idea about how strong the Higgs field should be
in the interior of the bound state, we may use the virial theorem.
According to the virial theorem, in the nonrelativistic
approximation which we use with a $1/r$ potential, the magnitude
of the potential energy has to be twice as big as the total
binding energy. Now we have precisely decided to adjust the
top-quark Yukawa coupling, so as to make the total binding energy
per constituent numerically equal to half its mass $m_t/2$. This
then means that the potential energy per top quark should be
$-m_t$ in the potential $\frac{1}{2} \sum_{j, \ j\neq i} V_{ij}$
felt by a constituent only feeling half the other 11 quarks.

We can understand that the change in energy of a top quark,
resulting from the reduction of the full Higgs field down from its
normal vacuum value to zero, would remove the mass and thus
correspond to a change by $-m_t$. So in the potential due to only
half of the constituents of the bound state we need just this
effect, meaning that the Higgs field should be zero (in the
approximation of ignoring the gluons) at the typical distance from
the center. Thus, taking into account that only half the
constituents are inside the average radius, we estimate that the
Higgs field at this average radius distance actually vanishes, $
\phi_h |_{at \ average \ pos.} = 0$. In the very most interior of
the bound state the effective Higgs mass is not so important,
since the distances are anyway small compared to the Compton wave
length of the Higgs. On the way out from the center, the effective
Higgs mass is small or even imaginary and only in the outskirts of
the bound state does it take on approximately its normal value.

Thus, at the average radius $<r>= 3/2 * r_0$, the potential energy
per top quark should be equal to $-m_t$, when we compensate the
total mass of the bound state and make it zero by letting the
binding energy be $m_t/2$ per constituent. This means that the
Higgs field is zero, $ \phi_h |_{at \ average \ pos.} = 0$, at
this average radius $<r>= 3/2 * r_0$.

Now the effective potential for the Higgs field has an inflection
point---i.e., second derivative zero---when its value is
$1/\sqrt{3}$= 0.58 times the value in the normal vacuum
$<\phi_h>_{normal} =v$. This inflection point value of the Higgs
field thus deviates from the normal vacuum expectation value of
the Higgs field by $(1-1/\sqrt{3})v$= $0.423v$, while the average
value of the Higgs field reached at $r =<r> =3/2* r_0$ deviates by
$v$ from the normal value. Since, in the first approximation, the
potential felt by the quark in the bound state goes down inversely
with the distance $r$ from the center, i.e., as $\propto 1/r$, the
inflection point is reached when $1/r$ has fallen by a factor of
$1/0.423 = 2.37$ compared to $1/<r> = 2/(3r_0)$. This means that
the inflection point is reached at the distance $r_{inflection} =
3/2 * 2.37 r_0 = 3.55r_0$.

\subsection{Correcting the Higgs field strength in the interior
due to the force being partly gluonic}

In Sec.~\ref{gluon} we calculated that approximately 1/3 of the
force responsible for the binding of the top and antitop quarks in
our bound state was due to gluonic rather than Higgs exchange.
Thus the binding energy from the Higgs exchange by itself should
only make up approximately $(2/3)^2 =4/9$ of the total binding
energy. Rather than having the potential energy per top quark
equal to $-m_t$ due to the Higgs field being zero at the average
distance from the center, as estimated above, we should instead
have that this average value of the Higgs field should be $ \phi_h
|_{at \ average \ pos.} = (1 - 4/9)v = 5v/9$. As we shall see in
Appendix G, we estimate that the field strength measured as the
deviation from the normal vacuum expectation value $v$, i.e.,
$-(\phi_h - v)$, reaches a value at the very center of the bound
state which is about 3/2 times the average deviation. Hence, when
the average of the Higgs field deviation is $4v/9$, this maximal
deviation---or the maximal field strength due to the top and
antitop quarks---will be $3/2 * 4v/9 = 2v/3$. Thus the actual
value of the Higgs field in the center of the bound state is
$(1-2/3)v = v/3$, meaning that it is one-third as strong as in the
usual vacuum. Hence the effective Higgs mass remains imaginary all
the way into the center, after we have passed deep enough into the
bound state for the value of the Higgs field to fall below its
value $v/\sqrt{3}$ at the inflection point.

The conclusion is that, closer to the center than the distance at
which the field $\phi_h$ has the strength $v/\sqrt{3}$
corresponding to the inflection point in the Higgs effective
potential, we have an imaginary effective Higgs mass. We will now
consider the real and imaginary effective Higgs mass regions
separately.

\subsection{The real Higgs mass region}

The only region in which we get a real effective Higgs mass is at
distances so far from the center that the value of the Higgs field
has risen above the inflection point value of $v/\sqrt{3}$. So let
us first consider the Higgs field in this region, where it is
numerically bigger than at the inflection point.

According to the above discussion, the average value of the Higgs
field in the region of the constituents of the bound state should
be $5v/9$. We take it that this value is reached at the average
position given by $r = <r> = r_0 *3/2$, where $r_0$ is the Bohr
radius. Since the Higgs effective potential has an inflection
point when the Higgs field takes on the value $\phi_h =
v/\sqrt{3}$, this must occur in the bound state when the distance
$r$ from the center has been increased relative to $<r>= 3/2* r_0$
by a factor $\frac{1-5/9}{1-1/\sqrt{3}}$. Thus, as one moves out
from the center of the bound state, the inflection point in the
Higgs effective potential is passed at the distance $ r =
r_{inflection}= \frac{1-5/9}{1-1/\sqrt{3}}*<r>$ = $1.052 *<r> =
1.58 r_0.$ With the probability distribution in $r$ taken to be
$\propto \exp(-2r/r_0)r^2dr$, the probability for a quark being
outside the distance characteristic of the inflection point
$r_{inflection}$ becomes crudely

\begin{eqnarray}
&&\frac{\int_{r_{inflection}}^{\infty}\exp(-\frac{2r}{r_0})r^2dr}
{\int_{0}^{\infty}\exp(-\frac{2r}{r_0})r^2dr}\\ & \approx &\exp\left(-\frac{2r_{inflection}}{r_0}\right)
\frac{\int_{r_{inflection}}^{\infty}\exp(\frac{-2(r-r_{inflection})}{r_0})r_{inflection}^2dr}
{\int_{0}^{\infty}\exp\left(-\frac{2r}{r_0}\right)r^2dr} \\
& = &\exp(-2r_{inflection}/r_0) \frac{\Gamma(1)r_{inflection}^2}{\Gamma(3)(r_0/2)^2}\\
& = &\exp(-3.15) *3.15^2/2\\ & =&21.2 \%.
\end{eqnarray}

Since the distance $r_{inflection}$ from the center to the
inflection point field value is only $5\%$ greater than the
average distance $ <r>$, the Higgs exchange Coulomb potential at
$r = r_{inflection}$ is only reduced by a factor
$<r>/r_{inflection} = 0.951$ compared to its value at the average
distance. So the effect of even an order of unity change of the
potential for $r > r_{inflection}$ (due to the Higgs mass effect)
could at most change the average of the overall binding potential
by the order of $0.951 *21.2 \% =20.2  \% $. At $r =
r_{inflection}$ the probability distribution $\propto
\exp(-2r/r_0)r^2dr$ of the quarks has a logarithmic derivative of
$(2/r_{inflection} - 2/r_0) = (2*0.951/<r> - 3/<r>) = -1.098/<r>$.
So the range, over which we have a significant part of the
probability, goes outside $r = r_{inflection}$ only by about a
distance of the order of $<r>/1.098$. In that range the effective
Higgs mass squared grows away from its starting value of zero at
$r_{inflection}$. By using a linear Taylor expansion in $\phi_h$,
we estimate that the effective Higgs mass squared reaches
$\frac{(1/1.098+ 1/0.9511)^{-1}}{1.098} = 0.464$ of its final
value at infinite distance. The value of the infinite distance
Higgs mass is the physical Higgs mass $m_h$, which we take to be
$115 \pm 50$ GeV in this article. So the effective Higgs mass in
the region of interest is $ m_{h\; eff} = (115\
\mbox{GeV})*\sqrt{0.464}= 78.3 $ GeV. The range over which this
Higgs mass is active is about $<r>/1.098$, so that the correction
factor, converting the Coulomb potential into a Yukawa potential,
becomes $\exp(- \frac{m_{h\; eff}* 3 r_0}{2*1.098}) =
\exp(-3\frac{m_{h\; eff}/(\sqrt{4/3}m_t)}{1.098 *2}) =
\exp(-3\frac{78.3/(172.6 \sqrt{4/3})}{2.196}) = \exp(-0.54)$.
However this $54 \%$ correction only applies to quarks at
distances $r > r_{inflection}$ from the center of the bound state.
So the percentwise correction to the total potential, due to the
Higgs mass in the real Higgs mass region, is $20.2 \%$ of $ 54 \%$
= $10.9 \%$. This effect gets doubled when calculating the binding
energy, because the radius varies with the strength of the
potential. However, since it is only for the Higgs part of the
potential, it should also be reduced by a factor 2/3. Finally then
we are interested in this paper in calculating the coupling
$g_t|_{phase \ transition}$, which is extracted from the fourth
root of the binding energy. So, at the end, this correction leads
to an {\em increase } in the phase transition coupling, needed to
get just zero mass for the bound state, by $ 2* \frac{2}{3} *
\frac{10.9}{4} \% =3.6 \% $.

\subsection{The imaginary Higgs mass region.}
As we have just seen the effective Higgs mass is imaginary in the
region of greatest relevance for the binding of the top quarks and
antitop quarks, namely, from $r =0$ out to where the Higgs field
takes on the inflection point value at the distance
$r_{inflection} = 1.58 r_0$ from the center. In Appendix G, we
crudely estimate the average effective Higgs mass squared, in this
region $0 < r < r_{inflection}$, to be $m_{h\; eff}^2 =
-m_h^2/12$.

The most important place to get effects from this effective
imaginary Higgs mass is from the very most central region out to
the average distance of the quarks and antiquarks feeling the
potential, which must crudely be at the distance $r= <r> =
\frac{3r_0}{2}$. The usual Yukawa potential having the form
$\propto \exp(-m_h r)/r$ should formally be replaced by a form
$\propto \exp(-i|m_{h\; eff}|r)/r$ in the imaginary effective
Higgs mass region. However it should be real in as far as the
Higgs field is ``real'' and, since the sign of the $i$ in the
exponent is ambiguous, we actually have to take \be \phi_h \propto
\cos(|m_{h \; eff}|r) /r \ee in the effective imaginary Higgs mass
region.

Actually it is not difficult to see that an expression of this
form obeys the Klein-Gordon equation with a tachyonic mass---i.e.,
$m_{h \; eff}^2 < 0$. Requiring the Higgs field to be given by the
Coulomb, i.e., massless, potential in the immediate neighborhood
of the particle emitting it, we also see that the only solution to
the Klein-Gordon equation with this boundary condition becomes the
cosine form just presented.

We take the averaged effect of the Higgs field on the binding to
be approximated by the  effect at the average distance $<r>$. This
means that the correction, due to the effective tachyonic Higgs
mass $m_{h \; eff}$ being imaginary, will become a factor
$\cos(|m_{h \; eff}|<r>)$ in the attractive Higgs exchange
potential between two (anti-)quarks. Since the latter is
proportional to $g_t^2$, this means that we effectively replace
$g_t^2$ by $g_t^2\cos(|m_{h \; eff}|<r>)$. That will in turn mean
that the $g_t$ value needed to achieve a certain condition for the
binding---in our case that we bind just so strongly as to make the
$6t + 6\overline{t}$ bound state massless---will have to be
increased by the factor $\sqrt{\cos(|m_{h \; eff}|<r>)}^{-1}$. In
other words
\begin{eqnarray}
&& g_t|_{phase \ transition} \rightarrow g_t|_{phase \ transition}/
\sqrt{\cos(|m_{h \; eff}|<r>)}\\
&= &  g_t|_{phase \ transition}/ \sqrt{\cos\left(\frac{m_{h}}{\sqrt{12}} *
\frac{3}{2} r_0 \right)}\\
&= & g_t|_{phase \ transition}/ \sqrt{\cos \left(\frac{3m_h }{8m_t} \right)}\\
&= & g_t|_{phase \ transition}/ \sqrt{\cos(0.250)}\\
&=& g_t|_{phase \ transition}* 1.016.
\end{eqnarray}
showing that  $g_t|_{phase \ transition}$ is increased by $1.6
\%$. Here we assumed the physical Higgs mass to be $m_h = 115$ GeV
and we used the crude estimate $|m_{h \; eff}| = m_h/\sqrt{12}$
from Appendix G. For the Bohr radius of the bound state, we took
$r_0 \approx (\sqrt{4/3}m_t)^{-1}$ from Appendix C. The average
radius is, of course, $<r> = 3/2*r_0$. Also we used the
experimental value \cite{mtop} of $m_t = 172.6$ GeV for the
top-quark mass.

Combining this with the  correction from the positive effective
Higgs mass region of $3.6 \%$, we get the total correction from
the Higgs mass not being zero to be a $ 1.6 \% + 3.6 \% = 5.2 \%$
increase in the value of $g_t$ needed for the phase transition.
We estimate a theoretical uncertainty of $\pm 2 \%$ in this result.
In order to take into account the $\pm 50$ GeV error in the Higgs
mass, we have repeated the above calculation for a Higgs mass of
165 GeV. We find the total correction in this case to be an
increase of $8.5 \%$ in the critical value of $g_t$. So we conclude
that, for a Higgs mass of $m_h = 115 \pm 50$ GeV, we obtain a
total correction of $5.2 \% \pm 3.3 \%$. Combining in quadrature
this $3.3 \%$ uncertainty arising from the error on the Higgs mass
with the estimated theoretical uncertainty on the calculation of
$2 \%$, we finally obtain the value $5.2 \% \pm 4 \%$ for the
increase in the value of $g_t$ needed for the phase transition.

\section{ s-channel exchanges}
\label{s6}
We only calculated the contributions to the binding energy from the $t$-channel
and $u$-channel exchanges above because:

a) These contributions are somewhat easier to calculate
in the Bohr atom approximation.

b) We believe that the $s$-channel contribution will be relatively
smaller due to the effect that, in an $s$-channel exchange, a
quark and an antiquark together with their associated binding
energy are virtually missing from the bound state. This leads to
an extra suppression of the binding energy from the $s$-channel
exchange.

In the present section we shall estimate the extra binding, due to
the $s$-channel exchange of both Higgses and gluons.

\subsection{ Crude channel symmetry estimation of s-channel contribution.}

First we shall make an estimate of the binding energy caused by
the $s$-channel effect---let us first consider just the Higgs
exchange---by thinking of an effective four quark interaction
term. We then compare the $s$-channel contribution to the
$t$-channel and $u$-channel contributions in such a formalism.

The plan is first to imagine a situation in which we could ignore
the masses of the quark and antiquark, interacting via the virtual
annihilation and recreation mechanism described by $s$-channel
scattering. The energy can then be chosen so that there would be a
symmetry between all three channels ($s$, $t$ and $u$), apart from
the selection rules. In this situation the dominant 4-momenta for
the quark (anti-)quark scattering comes from the 3-momenta arising
from the Heisenberg uncertainty in the momentum, which follows
from the geometrical extension of the wave function for the quarks
and antiquarks.

We may think of evaluating the binding energy, by taking the
expectation value of an operator corresponding to the Feynman
diagram for the Higgs exchange between a quark and an antiquark in
one of the three channels ($s$, $t$ or $u$). Such an expectation
value of a lowest order scattering operator should then be the
change in energy due to this interaction. Here we do not take into
account that, after the inclusion of some interaction, one should
also adjust the ground state wave function (e.g., the radius of
the bound state). We now imagine an artificial arrangement of
``small'' energies, replacing the ones due to the quark masses,
such that on the average the 4-momenta through the three channels
($s$, $t$ and $u$) are arranged to be the same\footnote{This same
value for the three quantities $s$, $t$, $u$ is of course not at
all consistent with the nonrelativistic situation, and strictly
speaking it is even in the unphysical region in the Mandelstam
diagram.}. Thereby the propagators in these different channels
will also be the same and thus the diagrams, when averaged, will
give the same numerical values, as long as they are not simply
forbidden by selection rules. This means that they would give
equal contributions to the binding energy. It is these imagined
momentum distribution configurations, which we want to use for
estimating the size of the $s$-channel
 contribution to the binding energy relative to that from the
$t$ channel. Then we must correct for the fact that these
artificially arranged 4-momenta get modified, when we instead take
the external 4-momenta to contain the quark masses in the
nonrelativistic situation. Furthermore, we must take into account
the effects of the lack of binding energy to the other quarks,
during the virtual time in which the pair of scattering quarks is
absent from the bound state.

Let us denote by $B$ the binding energy due to an allowed
$t$-channel exchange between two quarks, which is achieved without
changing the bound state wave function and is hence proportional
to $g_t^2$ rather than to $g_t^4$. Then this binding energy $B$ is
indeed the expectation value of the operator connected with the
$t$-channel exchange diagram for the scattering of the two quarks.

In the artificial situation proposed above, we arranged the energy
components of the four-momentum distributions for the quarks, so
that there was a symmetry between the three channels with respect
to these four-momentum distributions. This then means that not
only would the $u$-channel and $t$-channel interactions, counted
in the same way, lead to the same binding $B$, but even the $s$
channel would give the binding $B$ in the artificial situation.

Next we must estimate the change in the binding $B$, when we
include the correct rather than the artificial external 4-momenta.
The idea is that this makes no difference as far as the three
momentum is concerned. The major effect comes from the inclusion
of the correct mass energies and from the lack of binding to the
other quarks in the bound state during the $s$-channel quark
scattering. Thus there is no difference---at least in the
nonrelativistic approximation---to the 4-momenta in the $u$
channel and the $t$ channel. These $u$ and $t$ channels contribute
a binding energy $B$ by definition, and $B$ is not changed
relative to the artificial kinematical situation by including the
nonrelativistic masses into the energies. So we only need to get
the correct replacement for $B$ for the $s$-channel diagrams.

We now need to estimate the correction to the $s$-channel
propagator, by replacing the $s$-channel propagator with the
artificial four-momentum going through it by the one having the
correct four-momentum (mainly mass energy) going through it
instead. Now the artificial four-momentum going through the $s$
channel was precisely made up to be just the same as what goes
through the $t$ channel in the $t$-channel diagram. So really we
ask for the ratio of the $s$-channel propagator in the true
$s$-channel diagram to the $t$-channel propagator in the
$t$-channel diagram. Then we can correct the binding energy,
appropriately taken from the $t$ channel, by this factor and
thereby obtain the binding energy due to the corresponding
$s$-channel exchange term.

In order to perform this correction, we need to estimate not only
the binding $B$, which we have essentially already done in
previous sections, but also the average of the square of the
four-momentum going through the $t$-channel propagator.

\subsection{ Estimate of size of average t-channel momentum in propagator}

The average three momentum squared $\vec{q}^2 $ in the $t$-channel
propagator is achieved as the sum of the momentum distributions of
two of the Higgs emitting quarks---really the same quark before
and after the emission. Now we may easily estimate the expectation
value of the $\vec{p}^2$ distribution for the quark in the bound
state, using the virial theorem and the binding energy. In fact we
have from Eq.~(\ref{e18}) that \be m_t / 2 = \mbox{binding energy}
=-( V_{potential} + <\vec{p}^2/(2m_t)>). \ee As discussed in Sec.~5, 
it follows from the virial theorem that \be V_{potential}= -
2<\vec{p}^2/(2m_t)>, \ee which means that \be <\vec{p}^2/(2m_t)> =
m_t/2, \ee and hence \be <\vec{p}^2> = m_t^2. \label{e48} \ee So,
since the $t$-channel exchange goes between a quark to quark
transition vertex and another one, the probability distribution
for the momentum squared in the propagator should really be the
product of the distributions appearing from the two emissions. In
the Gaussian approximation the product distribution will have the
spread $<\vec{q}^2>$ obtained by adding the inverse $<\vec{q}^2>$'s,
i.e., $<\vec{q}^2>^{-1}$ for the two distributions multiplied.
These emission distributions in turn have, in Gaussian
approximation, the average of the $\vec{q}^2$ given as the sum of
that for the quark before and that after the emission. It is easy
to see that we then end up having the four or equivalently three
momentum squared in the $t$-channel propagator being the same as
the distribution of $\vec{p}$ for a single quark in the wave
function. In other words we obtain the propagator momentum squared
average \be <\vec{q}^2> = m_t^2. \ee

\subsection{Naive calculation with just the quark masses}

If we just calculate naively, according to the prescription
suggested, we should now simply insert the crude nonrelativistic
approximation $2m_t$ for the $s$-channel propagator four-momentum
value in the time direction, which dominates. This would mean a
decrease of the $s$-channel propagator by a factor 4 compared to
the one in the artificial situation or equivalently relative to
the $t$-channel one. This is only so simple because we ignore both
the Higgs mass and the lack of binding energy coming from the
quark-antiquark pair during their virtual annihilation time. This
result means that the binding energy due to the $s$ channel is
reduced from $B$ down to $B/4$.

In the following part of this section we shall correct this naive
s-channel binding energy expression $B/4$, by taking into account
the very strong interaction which the considered quarks,
annihilating into the Higgs, have with the other quarks in the
bound state. In our case, in which the binding cancels the mass
energy, of course this interaction must be very significant.
Although including such effects is in principle higher order and
really corresponds to calculating loop diagrams, we indeed need to
include them at least crudely. We shall perform these corrections
in a couple of steps:

1) We shall consider the Higgs relativistic Feyman propagator from
a nonrelativistic quantum mechanical second order perturbation
theory point of view, interpreting it to have two physically
different factors in the denominator; see Sec.~\ref{interprete}.

2) We shall take into account and estimate the extra energy
contribution accompanying the Higgs, due to the change in the
binding of quarks inside the bound state; see Sec.~\ref{intermediate}.

 \subsection{Comparing nonrelativistic perturbation with Feynman propagator}
 \label{interprete}

 It is well known that the relativistic Higgs-propagator is
 \be
 \mbox{prop}(p) = \frac{i}{p^2 - m^2} = \frac{i}{(p^0 - E(\vec{p}))(p^0 + E(\vec{p}))}.\label{Feynman}
 \ee
 This is made in a normalization of the Higgs field $\phi_H$
 given by the expression
 \be
 \int \phi_H^{\dagger} \stackrel{\leftrightarrow}{\partial}_0\phi_H d^3x
 =1.
 \label{normalization}\ee
This normalization deviates from the simple nonrelativistic one:
\be
 \int \phi_{H \ nr}^{\dagger} \phi_{H \ nr}d^3x =1.
 \ee
For approximate energy eigenstates with energy $E_{Higgs}$, this
implies the following relationship between the relativistically
and nonrelativistically normalized fields:
 \be
 \phi_H = \frac{1}{\sqrt{2E_{Higgs}}} \phi_{H \ nr}. \label{e58}
 \ee

 The interaction energy density  of the Yukawa term in the Lagrangian
 becomes
 \be
 {\cal H}_{Yukawa} = -g_t( \bar{\psi}_{tR}\phi_H^{\dagger} \psi_{tbL} + H.c.)
 \ee
 [see Eq.~(\ref{yH}) in Appendix A
for notation], where the field $\phi_H$ is the relativistically
normalized field. Thus, in nonrelativistic notation, this
Hamiltonian density would rather look like:
 \be
 {\cal H}_{Yukawa} = -\frac{g_t}{\sqrt{2 E_{Higgs}}}( \bar{\psi}_{tR}\phi_{H \ nr}^{\dagger}
 \psi_{tbL} + H.c.). \label{nry}
 \ee

 Now, according to usual nonrelativistic second order perturbation theory,
 one has the correction to say the energy of the ground state $ |gs>$ from this second order
 effect:
 \be
 <gs| \int {\cal H}_{Yukawa} d^3x |Higgs><Higgs| \int {\cal H}_{Yukawa} d^3x |gs> / ( E_{Higgs} - E_{gs}).
 \label{nrper}
 \ee
For example, say we wanted to consider the change in energy of a
quark-antiquark pair due to $s$-channel Higgs exchange, then
$E_{gs}$ would be the energy of the unperturbed pair and
$E_{Higgs}$ would be $E(\vec{p}) = \sqrt{m_h^2 + \vec{p}^2}$.
 In Eq.~(\ref{Feynman}) the denominator factor $p^0 - E(\vec{p})$ is thus to be
 identified with the denominator in the nonrelativistic perturbation correction (\ref{nrper}),
 i.e.,
 \be
 p^0 - E(\vec{p}) =-( E_{Higgs} - E_{gs}).
 \ee
In the nonrelativistic notation, using (\ref{nry}), and for $p^0$
close to the on-shell energy of the Higgs, the matrix elements in
(\ref{nrper}) each contain an extra denominator
$\sqrt{2E_{Higgs}}$ compared to the relativistic notation, which
we can transfer to the propagator. In this way we get a propagator
to be used, together with the formal relativistic notation matrix
element, without such a denominator, \begin{eqnarray}
\frac{-i}{E_{Higgs} - E_{gs}} * \frac{1}{\sqrt{2
E_{Higgs}}{\sqrt{2 E_{Higgs}}}} = \frac{i}{(p^0 - E(\vec{p}))(2
E_{Higgs})} &\approx&  \frac{i}{(p^0 - E(\vec{p}))(p^0 +
E(\vec{p}))} \nonumber\\ &=& \mbox{prop}(p). \end{eqnarray}

It will be important for us to use this physical interpretation of
the two different factors in the denominator of the relativistic
propagator, when in the next section we shall take into account
the very strong interaction of the quarks, which annihilate into
the Higgs, with the rest of the quarks in the bound state.

\subsection{ Extra energy in the intermediate state}
\label{intermediate}

The important effect of the strong interaction, between the two
annihilating quarks and the other quarks, is that the energy of
the remaining 10 quarks (really 5 quarks and 5 antiquarks) may be
changed drastically by the absence of the annihilated quarks. This
change in energy means that the energy of the intermediate
state---which is talked about here as the Higgs state---is
actually shifted relative to the Higgs energy proper to an
effective Higgs energy including this interaction energy change.

When we want to include the missing binding energy of the
annihilated pair together with them into the calculation, one
should strictly speaking consider the whole process described by
an effective loop Feynman diagram, in which the bound state of 12
particles (the t ball) is split up into a Higgs and a ``core''
consisting of a bound state of 10 constituents. The loop vertex
should then really be a description of the annihilation process
coupling to the emitted Higgs. If we indeed went into the details
of the estimation of such a loop, we would have to integrate over
a loop energy, $p^0$. The integrand would have poles coming from
both the Higgs and the core (i.e., the 10-constituent bound
state). In fact we propose to look at the contributions from near
these poles as two terms to be calculated separately. In order to
avoid going into the details of loops, we shall however make
another presentation, in which we instead only talk about tree
diagrams. The price, however, is that now we must vary what state
we take as the background (or one could say the vacuum), in
evaluating what we believe would be the same contributions that
come from the different poles in the loop formulation.

We can indeed consider the following two points of view, with
respect to the vacuum for our problem:

1) We choose the ``vacuum'' to be the full 12 component bound
state with an extra Higgs present, in a state with the spatial
Higgs momentum distribution which we estimate couples to the
annihilation. The initial state, consisting of the 12-constituent
bound state without any extra Higgs, now has an energy below that
of the vacuum, because of its {\em lacking} Higgs. That is to say
the initial state has an energy $-E_{Higgs}$, where $E_{Higgs}$ is
the energy of the Higgs in the vacuum. So we think of this as the
initiating $q \bar{q}$ pair having the initial energy
$-E_{Higgs}$.

Now the process is that a $t \bar{t}$ pair annihilates to become a
hole (really a double hole) in the vacuum, because the vacuum
should have 12 constituents and after the annihilation there are
only 10 left. So they really form a virtual $s$-channel hole. This
hole represents that we have the 10-constituent bound state
instead of the 12-constituent one. As we shall see in Appendix H,
the mass difference between these bound states is $m_{10} - m_{12}
\approx 950$ GeV. So the hole must be counted as having the energy
$950$ GeV. Since we start with a state with energy $-E_{Higgs}$,
taken to be of the order $ - m_t$ because the momentum is of that
order, it means that the (double) hole must be strongly off shell.

In the relativistic notation we formally get a propagator with a
denominator of the order of $950$ GeV to the second power, which
means that we assume it to be smaller than the corresponding
object in the $t$ channel by a factor $(950\ \mbox{GeV}/m_t)^2=
(950/172)^2 = 5.5 ^2 =30.5 $. But now, if we want to write the
diagram in terms of the nonrelativistic vertex form (see 
Sec.~\ref{interprete}), there is in this form a factor
$1/\sqrt{2E_{hole}(\vec{p})}$ for each of the two vertices that
must be extracted to get the vertex in the nonrelativistic
formulation (\ref{e58}). Using the nonrelativistically normalized
field $\phi_{hole \ nr}$, we would expect the transition matrix
element between the 12-constituent bound state and the
10-constituent one to be a rather simple overlap giving just unity
in first approximation---of ignoring, for example, the difference
in radii. Thus one of the two factors of 5.5 is used up by the
factor $1/2E_{hole}$. So, as we think of varying the ``big''
number 5.5, we only get the $s$-channel contribution suppressed by
one factor of 5.5. Since we assume that for the hole energy of the
order of $m_t$ only we would have gotten the same as in the
$t$-channel case, this means that the suppression of the
$s$-channel contribution is by the factor 5.5. However, we did not
include the kinetic energy resulting from  the spatial momentum
being of the order of $m_t$, as given by Eq.~(\ref{e48}). This
means that the true suppression factor is rather $\sqrt{5.5^2 +1}
= 5.6$.

2) In this case we consider a vacuum which is simply the bound
state with 10 constituents and we consider the Higgs to be the
$s$-channel particle. Then the initial state has all the extra
binding energy of the 12-constituent state compared to the
10-constituent one. That is to say now the initial energy is
$-950$ GeV. In the Higgs propagator the factors in the denominator
are of this order of magnitude, but we cannot absorb such strong
suppression from even one of them by crudely identifying it with
the factors $1/\sqrt{2 E_{Higgs}}$ contained in the Higgs
vertices, because the $E_{Higgs} $ in the latter is given by the
Higgs momentum and mass and these quantities in our model never
reach more than about $m_t$. So, in this case 2, we indeed get a
very small contribution only of order $ 1/5.5^2 = 1/30.5$ compared
to the $t$ channel or rather $1/5.6^2 = 1/31.4$, when we include
the spatial momentum.

As we shall see in the following section, these two different
tree-diagram estimates should really be {\em added}. In other
words, the full $s$-channel contribution is suppressed, relative
to the analogous $t$-channel term, by a suppression factor equal
to the sum of the two above computed suppression factors: \be
``\hbox{suppression factor}'' = 1/5.6 + 1/5.6^2\approx
\frac{1}{5.6(1-1/5.6)} \approx 1/4.6. \ee Thus we shall calculate
an $s$-channel contribution by first evaluating the coupling and
combinatorial factors and then dividing the result by 4.6. We
shall do this for the Higgs exchange in Sec.~\ref{gteval} and for
the gluons in Sec.~\ref{gleval}.

\subsection{Arguing for adding the two terms}
\label{adding}

From the above discussion it may not be clear what we have to do
with the two different results obtained under the points of view 1
and 2, respectively. We want to argue here that we should indeed
add these two contributions. However, for this purpose, it is best
to think of doing the calculation as a loop correction. Then we
look at the correction to the binding energy as the result of the
virtual split up of the 12-constituent bound state (the t ball)
into a Higgs particle and the bound state consisting of only 10
constituents. This means that it is truly a self-energy diagram in
an effective field theory (with the various bound states as
particles giving Feynman rules together with, e.g., the Higgs),
which corrects the mass of the 12-constituent bound state.

When we formulate the mass correction in this loop way, we end up
with a loop four-momentum $q$ over which to integrate. Let us now
think of the performance of the integral over the energy component
 $q^0$ of this loop four-momentum $q$: For fixed values
 of the spatial components of the loop four-momentum $\vec{q}$
 the integrand is (basically) a product of two propagators,
 namely, one for the 10-constituent bound state and one
 for the Higgs. It therefore gets poles whenever one
 of these two particles is on shell. We imagine
 to approximate the whole loop integral by the sum over contributions from
 the neighborhood of these poles. Actually it is not difficult
 to see that, by an appropriate closing and deformation
 of the contour, you can prove that the loop integral
 over the $q^0$-dummy variable gives a sum over the pole residues
 (divided by $2\pi$). Now the point is that these
 pole contributions can indeed be identified with the results
 from the formal tree diagrams just discussed under points 1 and 2.
 In fact the contribution from the Higgs-propagator pole
 (for positive Higgs energy) in the loop integrand gives
 us the formal tree diagram corresponding to the on-shell
 Higgs being considered part of the vacuum. The propagator
 in this formal tree diagram
 corresponds to the hole in the other 12-constituent part of the
 vacuum, so that it is really the propagator for the 10-constituent
 particle that lies under the hole. Thus this contribution
 from the pole of the Higgs propagator in the loop corresponds to
 case 1 above.
 Similarly the residue contribution from the pole
 of the 10-constituent bound state propagator in the loop integrand gives the
 contribution in which this 10-constituent bound state
 is identified with the vacuum. This is case 2 above.

 Since we have now identified the two tree-diagram contributions
 from the previous section as being
 two contributions coming out of the same loop integral,
 we see that these contributions to shifting
 the mass of the 12-constituent bound state must be {\em added}.

\subsection{Finding the s-channel Higgs correction to $g_t$}\label{gteval}

A certain quark in the bound state can only annihilate together
with the antiquark having just the compensating color and spin. So
there is among the antiquarks only one that can annihilate with a
given quark into the Higgs. This means that the factor 11,
corresponding to the number of quarks or antiquarks that can
interact via $t$-channel exchange with a given quark, gets
replaced by 1 for the $s$-channel exchange. In Sec.~\ref{2p1} we
saw that, for $u$-channel exchange, we had to replace this factor
of 11 by 5. So, by including the $u$ channel, the interaction of a
quark by Higgs exchange has a combined strength of coupling to the
other constituents as if there were 16 of them coupling by only
$t$-channel exchange. Thus, if the strength of the $s$-channel
coupling, when allowed, had been just the same as for the $t$
channel, meaning just $B$, then the coupling strength of the $s$
channel would have made up $\frac{1}{16}$ that of the $t$- plus
$u$-channel Higgs exchange. Now these coupling strengths or
scattering amplitudes are  proportional to $g_t^2$. Thus, if the
$s$-channel Higgs exchange results in a $1/16 = 6.25 \%$ increase
in the scattering amplitude, then we should decrease the
previously predicted critical coupling $g_t|_{phase \ transition}$
by $\frac{1}{2} * \frac{1}{16} = 3.125 \%$. But now, as we
estimated above in Sec.~\ref{intermediate}, the $s$-channel
propagator has to be suppressed by a factor of 4.6. This means
then that, if we totally ignore the gluons, the percentwise
decrease of the previously calculated critical $g_t|_{phase \
transition}$ would be $ \frac{1}{32* 4.6} = \frac{3.125}{4.6}\%  =
0.68 \%$. Since the gluon contribution to the potential does not
depend on $g_t^2$, the correction of including the $s$-channel
Higgs exchange will change the critical $g_t$ = $ g_t|_{phase \
transition} $ downward by $0.68 \%$, i.e., \be \Delta \ln
g_t|_{phase \ transition} = - 0.68\% \ \mbox{(from $s$-channel
Higgs)}. \ee

\subsection{The gluon s-channel correction}\label{gleval}

Next we shall consider the change in the scattering amplitude, or
equivalently the potential, from the exchange of gluons in the $s$
channel. Each quark can interact by annihilating into a gluon with
any one of the antiquarks, except when they form a color singlet
together. We can take care of the latter exception by including a
correction factor 8/9 in the scattering amplitude. Apart from this
exception, we have interaction between all quarks with antiquarks,
while neither quark and quark nor antiquark and antiquark can
annihilate into gluons. So one quark can interact via the $s$
channel with 6 antiquarks. Thus we can estimate the strength of
the $s$-channel gluon exchange, counted in amplitude or potential,
as being 6/16 times as strong as the $u$- plus $t$-channel Higgs
exchange, provided we replace the Higgs coupling $g_t^2/2 =
(0.935)^2/2 = 0.437$ by the equivalent gluon coupling
$e_{t\overline{t}}^2 =  1.83$ [see Eq.~(\ref{ett})]. This
replacement gives an increase in strength by a factor of 4.2. We
must also remember to include the correction factor of 8/9. So
finally we get the $s$-channel gluon exchange binding amplitude to
be given relative to the combined $u$- and $t$-channel Higgs
exchange by \be \frac{\mbox{Gluon $s$-channel}}{\mbox{Higgs
$t$-channel   + $u$-channel}} = \frac{6}{16}* \frac{4.2}{4.6}
*\frac{8}{9} = 0.304. \ee Here we also included the suppression
factor of 4.6 from Sec.~\ref{intermediate} for the $s$ channel.
Now, since the amplitude in which we calculated the correction is
proportional to $g_t^2$, we obtain a backcorrection in $g_t^2$ of
$30.4 \%$, meaning that $g_t^2 $ after the correction has to fill
in the same as $g_t^2 $ before with $30.4 \%$ subtracted, i.e.,
$g_t^2|_{corrected} = (1-30.4\%)g_t^2|_{before}$. So the
correction to $g_t|_{phase \ transition}$ due to $s$-channel gluon
exchange is downward by $-\ln (1-0.304)/2$ = $18.1 \%$, i.e., \be
\Delta \ln g_t|_{phase \ transition} = -18.1 \% \ \mbox{(from
$s$-channel gluons)}. \ee

\subsection{s-channel summary}
Summarizing we obtain  \be \Delta \ln g_t|_{phase \ transition} =
-18.1\% - 0.68 \% = -18.8\% \ \mbox{(from full $s$-channel)} \ee
for the total correction from the $s$ channel counted
logarithmically.

\section{Top mass field dependence corrections}\label{s7}

If we could be allowed to use the masses $m_i$ for the constituent
particles undisturbed by the Higgs field having  different values
in different places in the interior of the bound state, then the
expression (\ref{eq25}) for the infinite momentum frame energy,
derived in Appendix B, would lead to the expansion for
$m_{bound}^2$ in Eq.~(\ref{e10}) with $E_{binding}$ being just the
nonrelativistic expression formally, even if this binding is big
compared to the mass terms. In this sense the infinite momentum
frame expansion justifies the formal nonrelativistic calculation,
provided we take the former to mean the expansion of the mass
squared being extrapolated without higher order terms.

Now, however, the masses occurring in this formula are supposedly
changed, due to the average Higgs field in the interior of the
bound state being smaller than in the outside. Such a change of
the effective top-quark mass will naturally change the mass of the
bound state and, at first, it looks like we should include a
correction for this effect.

However, we see that in the approximation of the masses all being
scaled by the same factor, due to the averaged Higgs field in the
region where they are on the average, the whole bound state mass
squared will simply be scaled by the square of this factor. This
is simply a consequence of a dimensional argument, since the mass
is the only dimensional quantity entering the calculation. The
quantity $p_z$ is, namely, only a formal going to infinity
quantity.

Now, however, the quantity we are truly after is just the $g_t$
value $g_t|_{phase \ transition}$ at which the mass squared of the
bound state becomes zero. That is, however, a dimensionless
quantity being asked for, and that {\em cannot depend on the value
of the single mass scale quantity, the average mass}. Thus there
should be no change in our phase transition coupling prediction
due to such an effective mass change, provided we can count it as
being by the same factor crudely all over the inside of the bound
state. Thus actually, in the first approximation, no corrections
are needed. This means $0 \%$ correction to first approximation.

\subsection{Next order correction in effective mass variation with the field}

Now, however, the approximation in which the effective mass inside
the bound state should be just the same all over in space is not
so terribly good. Rather we must take into account that, for a
quark being in the deep interior of the bound state, the effective
mass is smaller than for one being farther out in the outskirts of
the bound state.

In order to correct for this {\em variation} of the effective
mass, we imagine to have calculated the average mass $m_{av}$
corresponding to the average Higgs field felt by the top quark.
Then we may write the true space-dependent effective mass as \be
m(\vec{r}) = m_{av} + \Delta m(\vec{r}). \ee

In Sec.~\ref{sec5} about the Higgs mass correction, we found that
even at the center of the bound state the Higgs field was
estimated to be 1/3 of its faraway value, i.e., $v/3$, while on
the average with respect to the constituent distribution it was
$5v/9$. In the classical approximation the constituent can only
reach out to the distance $r$ where the kinetic energy becomes
zero. Using the virial theorem, this corresponds to where the
potential has fallen to numerically half the value at the average
distance $<r>$. At this classical upper limit for the radial
distance $r$, the field $\phi_h $ must be in the middle between
the faraway value $v$ and $5v/9$. Hence, at the classical boundary
for the constituents, the Higgs field is $7v/9$. This already
gives us an estimate of the fluctuation in the effective mass \be
\frac{|\Delta m|}{m_{av}} < \frac{5/9 -1/3}{5/9} = \frac{2}{5}
=0.4 \ee or \be \frac{|\Delta m|}{m_{av}} < \frac{7/9 -5/9}{5/9} =
\frac{2}{5} =0.4. \ee {\em A priori} these coincident estimates
are even overestimates and should be reduced by considering a flat
interval distribution between $v/3$ and $7v/9$. Then, using
$\frac{\int_{-1}^1x^2dx} {\int_{-1}^1dx} = 1/3$, we obtain a
reduction factor of $1/\sqrt{3}$, which gives \be
\frac{\sqrt{<\Delta m^2>}}{<m>}= \frac{|\Delta m|}{m_{av}} \approx
\frac{5/9 -1/3}{5/9
* \sqrt{3}} =\frac{2}{5\sqrt{3}}= 0.23. \label{Dmm} \ee

\subsection{An alternative mass fluctuation estimate}\label{alternative}

Another estimate of this ``fluctuation'' in the effective quark
mass is gotten by using the fact that the relative spread in the
radial distance is \be \frac{\sqrt{<r^2 - <r>^2>}}{<r>} =
\frac{1}{\sqrt{3}}, \ee which in turn implies a spread in the
potential energy \be \frac{\sqrt{(<V-<V>)^2}}{<V>} \approx
\frac{1}{\sqrt{3}}. \ee Since $V \propto m_t - m(\vec{r}) $, this
means that \be \frac{\sqrt{<(m_{av} - m(\vec{r}))^2>}}{m_t -
<m(\vec{r})>} \approx \frac{1}{\sqrt{3}}, \ee and thus, using
$m_{av} = \frac{5}{9} m_t$, we get \be \label{f046}
\frac{\sqrt{<\Delta m(\vec{r})^2>}}{m_{av}} \approx
\frac{4}{5\sqrt{3}} = 0.46. \ee But here we did not take into
account the flattening off of the potential in the center
discussed in Appendix G, which we used in the first estimate.

Instead of arguing via first estimating the fluctuation in the distance
and then calculating as if this fluctuation were small, we can
directly calculate the fluctuation in $1/r$ or equivalently the
Coulomb potential. In this case we get a value of 4/5, which is
$\sqrt{3}$ times bigger than (\ref{f046}).

\subsection{Taylor expanding in the mass}

In the nonrelativistic looking condition for the binding energy
per particle just being equal to $m(\vec{r})/2$, given in
Eq.~(\ref{e18}), the only $m(\vec{r})$-dependent term with nonzero
second derivative with respect to $m(\vec{r})$ is the kinetic
energy term $\frac{\vec{p}^2}{2m}$. This term has the second
derivative \be \frac{\partial^2 (\frac{\vec{p}^2}{2m})}{\partial
m^2} = \frac{\vec{p}^2}{m^3}. \ee Provided that the average of the
square of the $\Delta m(\vec{r})$ is as given by (\ref{Dmm}),
i.e., $<(\Delta m)^2> = 0.23^2 *m_{av}^2$ = $0.053 m_{av}^2$, we
obtain an effective replacement for the kinetic term:
 \be
 \frac{\vec{p}^2}{2m} \longrightarrow \frac{\vec{p}^2}{2m} + \frac{1}{2}\frac{\vec{p}^2}{m^3} * <(\Delta m)^2>=
 \frac{\vec{p}^2}{2m} + \frac{1}{2}\frac{\vec{p}^2}{m^3} * 0.053 m_{av}^2 =
 \frac{\vec{p}^2}{2m}( 1 + 0.016)
 \ee
\subsection{Correction to $g_t|_{phase \ transition}$ from kinetic term mass fluctuation change}

The change of the kinetic term effectively due to the mass {\em
variation} by the factor $(1 + 0.016)$ means that, in
Eq.~(\ref{eigen}) for the Hamiltonian in Appendix B, we have
replaced the top-quark mass by a value $(1+0.016)^{-1}$ times as
big. Thus the binding energy resulting from use of the modified
version of this expression will, for dimensional reasons, be
$(1+0.016)$ times smaller than the usual Rydberg (\ref{binding}).
To compensate for this decrease in the binding energy, the fourth
power of $g_t$ to which the Rydberg is proportional must be
increased by $1.6\%$. Thus this correction, due to the variation
of the effective mass over the bound state volume, to our critical
Yukawa coupling prediction is that we increase the prediction by
$0.4\%$.

Had we, instead of (\ref{Dmm}), used the alternative estimate
(\ref{f046}) for the variation of the effective mass, we would
have gotten a $4$ times bigger value for $<(\Delta m)^2>$. This
would, in turn, mean an increase in the value for the predicted
critical Yukawa coupling  of $0.4 \% * 4 = 1.6 \%$. Had we used
the even bigger estimate at the end of Sec.~7.2 for the
fluctuation in the mass \be \frac{\sqrt{<\Delta
m(\vec{r})^2>}}{m_{av}} \approx 4/5 = 0.80, \ee
 we would have gotten an increase of $4.8 \%$ in our predicted
critical Yukawa coupling.

 Since the bigger estimates of the correction correspond to using an unsmoothed
 potential even near the center of the bound state, they
 are probably less reliable. So we
 have a bit more confidence in the $0.4 \%$ estimate;
 but let us take $\Delta \ln g_t|_{phase \
transition} = 2\% \pm 3 \%$ as a reasonable average.

\section{Finite speed of Higgs exchange} \label{finitespeed}

In the non-relativistic calculations which we used, we took
the interactions to be instantaneous and ignored the fact
that the Higgs or gluon being exchanged between a couple of quarks
or antiquarks after all only travels with the speed of light.
Under such conditions the only Feynman digrams for $t$-channel
exchange are the diagrams in which Higgses or gluons are
exchanged one after the other. However, diagrams, in which a couple
of quarks among our 12 interact by an exchange of two
Higgses propagating simultaneously, are ignored in this
approximation. By this we mean that a diagram in which the two
Higgs propagators cross each other, when being exchanged, is what
is ignored in the nonrelativistic approximation we used. We
should however, to higher accuracy, include such possible effects
of the emission and the absorption of the exchanged Higgs
not being quite simultaneous.

We shall do this crudely here, by estimating the fluctuation
caused by this effect in the distance $r$ between the interacting
quarks to be used in the potential (\ref{f1}): By the virial
theorem, we have that the kinetic energy of a quark in its motion
in the potential equals minus one-half of the potential energy
and, thus, is just equal to the binding energy numerically. Since
the binding energy, in the critical case which we look for, is
$m_t/2$, we obtain on the average \be
<\frac{\vec{p}^2}{2m_t}> = m_t/2. \ee This implies that for a
single component of the momentum---e.g., the component along the
line connecting the interacting quarks---we have on the average
\be <p_x^2> = m_t^2/3. \ee This implies a velocity component with
a spread, due essentially to quantum fluctuations, of \be <v_x^2>
= 1/3. \ee This in turn implies that the effective distance $r$ to
be used in the potentials such as (\ref{f1}) actually fluctuates
by $\sqrt{1/3} * 100 \%$ = 57$\%$. Now the second derivatives of
the potentials such as (\ref{f1}) are of the form \be
\frac{d^2V_{t-channel \ Higgs}}{dr^2} = -\frac{2*g_t^2/2}{4\pi
r^3}. \ee Hence, by Taylor expanding the potential at the
Higgs-delay corrected distance $r_{dc}$ around the first or
nonrelativistic approximation value $r_{nr}$ for the distance
between the interacting quarks, we get the fluctuation corrected
effective potential to be:
\begin{eqnarray}
V_{eff}(r_{nr})& = &<V_{t-channel \ Higgs}(r_{nr}) +\frac{dV_{t-channel \ Higgs}(r_{nr})}{dr_{nr}} (r_{dc} - r_{nr})\\
&&+\frac{1}{2} \frac{d^2 V_{t-channel \ Higgs}(r_{nr} )}{dr_{nr}^2} (r_{dc} - r_{nr})^2 +...>\\
&=&  V_{t-channel \ Higgs}(r_{nr})
+\frac{1}{2} \frac{d^2 V_{t-channel \ Higgs}(r_{nr} )}{dr_{nr}^2} r_{nr}^2/3 +...\\
& = & (1 + 1/3) V_{t-channel \ Higgs}(r_{nr}) + ... \label{delay}
\end{eqnarray}

Thus, at the end, we get that this effect of the delay of the
propagation of the (in first approximation) infinite speed Higgs
exchange causes an effective spread in the distance $r_{dc}$ to be
used for evaluating the potential. This causes an effective
increase in the potential by a factor of $1+1/3 =4/3$, in our
situation corresponding to the critical case of a zero mass bound
state. In turn this means that the coupling $g_t$
needed to provide this critical mass zero bound state should be
corrected, by reducing it by the square root of this factor of
4/3. This means that, instead of Eq.~(\ref{f17}), we get
the same equation but with the factor $0.0929$ replaced by a
number which is $(4/3)^2$ times bigger. So the equation now reads
\be 0= 1 -0.0929*(4/3)^2*(0.456 +g_t^2)^2. \label{f17n} \ee Thus
we obtain the following value for the critical $g_t$, evaluated
using the delay corrected effective potential (\ref{delay}):
\begin{equation}
\label{gtphasen} g_t|_{phase \ transition}\simeq
\sqrt{\sqrt{1/(0.0929*(4/3)^2)} - 0.456} = 1.42.
\end{equation}
Compared to the previous value for the critical $g_t$ at this
stage---before even the introduction of the $b$ quark correction
of Sec.~\ref{sec3}---this is a downward correction given
logarithmically percentwise by \be \Delta \ln g_t|_{phase \
transition} = \ln\frac{1.42}{1.68} = - 17.2 \%. \ee

\section{Many body correction}\label{s9}

Clearly the calculation made as if all the other quarks or antiquarks than the one considered were
sitting in just one point cannot be correct; so we have in principle to make calculations on
the system of the 12 constituents as a true many body system.

Here we shall do this in a rather crude way, only thinking of an
ansatz in which the constituents are described by a factorizable
wave function, meaning that it is a product of a wave function for
each constituent independently of the other ones. Then it is
obvious that the spread in the distance between a couple of
constituents will be just $\sqrt{2}$ times bigger than that of the
independent particle distributions. In turn this means that, to
the extent that the expectation value of the momentum squared is
given by---or at least varies as---the Heisenberg uncertainty
relation, the independent $<\vec{p}^2>$ will be twice that of the
relative motion of one pair. This change will function as if the
mass in the kinetic term were, for the many particle description,
smaller by a factor 2 than in the starting relative position
description. For dimensional reasons such a diminishing of the
mass by a factor 2 would also diminish the resulting binding
energy by this factor 2. In turn that would mean that we
should correct our critical coupling upwards by a factor of
$2^{1/4}$. This means a logarithmic percentwise increase of
$100\%
* \ln(2)/4 = 17.3 \%$.

The many body corrections we are studying here reflect the fact
that the calculation of the quark contributions to the binding, as
if the individual pairs of top or antitop quarks could distribute
themselves so as to minimize the energy of just that pair, cannot
be quite correct. If two of the constituents are not essentially
at the same site, it is impossible for a third one to be
very close to both. In Appendix I we have illustrated this problem
of the impossibility of having all the pairs have their optimized
relative distance distribution, by using a Gaussian ansatz
factorizable wave function for the whole bound state. Indeed the
factor of 2 correction discussed in the previous section is
essentially realized, but more precisely a factor of 2.16 is
obtained (see Eq.~(\ref{216factor})). This corresponds to an
upward correction of $100\% * \ln(2.16)/4 = 19.3 \%$ in our
critical coupling $g_t|_{phase \ transition}$, which provides an
upper bound for the many body correction considered. In fact one
could {\em a priori} very easily imagine that, by making a more
complicated ansatz wave function, we could enhance the probability
for the individual pairs having a small relative distance. In this
way we could make the distribution between the constituents in a
pair approach more closely to the ideal ground state distribution
for a two particle system. Certainly we must expect that the true
wave function for the bound state system must have gone a bit in
this direction compared to the ansatz in Appendix I.

It may be best to think about this effect, of somehow getting the
wave function improved to cluster the constituents more on the
short distances, as an antiscreening that could even be
approximately described by a ``dielectric'' constant for the
medium of constituents conceived as a material. With such a
dielectric constant $\epsilon$, the potential around a
constituent is modified from the usual $\frac{g_t/\sqrt{2}}{4\pi
r}$ form to  $\frac{g_t/\sqrt{2}}{4\pi \epsilon r}$.

Let us now attempt to estimate an effective $1/\epsilon$ correction to use (on the potential)
as a function of $r$.
Such a correction factor $1/\epsilon$ would correct the
quantity $g_t^2$ in our expression for the potential, which we might think of as being
in an effective distance $r$ dependent way.

All our earlier calculations before this section were made without
any ``many body'' correction and assumed the absolutely most 
well-arranged relative distributions for all the pairs. However it is
not possible to realize such a distribution for all the quarks
simultaneously and thus these previous calculations provide an
upper limit to the correction factor. It is therefore impossible
that the correction factor $1/\epsilon$ could be more than a
factor $\sqrt{2.16}$.

When we then think of the correction factor as dependent on an
effective distance $r$, we must imagine a function of this $r$
taking values between $1$ and $\sqrt{2.16}$. It is clear that, for
the distance $r$ going to zero, it is hopeless to organize
clustering and the correction factor must go to $1$ there. Also at
$r \rightarrow \infty$, where we think of a constituent isolated
from the rest, there are essentially no particles to cluster with
and no further clustering is possible. In practice the rest of the
particles are already clustered in this case. So the further
correction factor can only be $1$ in this limit too. In the
intermediate region in $r$, you would however expect some further
clustering to take place compared to that of the ansatz wave
function in Appendix I. So let us now assume that the correction
factor as a function of $r$ is reasonably smooth, say basically a
second order polynomial, in the range of any significant
population of the constituent distance $r$.

The maximal possible modification of our above correction of $19.3
\%$ could now only be achieved by having the maximal correction
factor $1/\epsilon = \sqrt{2.16}$ around the typical or most
likely distance, i.e., around $r = <r>$. But then the correction
factor must also reach $1/\epsilon =1$ as $r$ goes to zero or to
infinity effectively. Roughly this must mean that, for the tails
of the distribution to both sides, we get the correction factor
$1$ rather than the $\sqrt{2.16}$. Let us very crudely estimate
that, averaged over the distribution, we get the mean between the
two values $1$ and $\sqrt{2.16}$. That would mean that we would
get $g_t^2$ replaced by $g_t^2 * (1 + \sqrt{2.16})/2$ or 
$g_t^2 * (2.16)^{1/4}$ using a geometrical mean instead. In order
to compensate for that, we would need to decrease the
critical coupling $g_t|_{phase \ transition}$ by a factor of
$(2.16)^{1/8}$. This means a decrease of the critical $g_t$
prediction from our model by $10.1 \%$.

Together with the $19.3 \%$ increase, this ``backcorrection''
means that we would end up with a $19.3 \% - 10.1 \%$ = $9.2 \%$
correction. It seems reasonable to consider this latter value,
i.e., a $9.2 \% $ increase of $g_t|_{phase \ transition }$, as a
lower bound for the many body correction. Therefore, crudely, we
might present the result of this rather big many body correction
as an increase of the predicted critical coupling by $ (19.3 +
9.2)\%/2 \pm 9.2\%/2 = 14.2 \% \pm 4.6 \%$.

\section{The SU(2) part of $Z^0$ and $W$ exchange effects}\label{s10}

We expect the effect of exchanging the time-components or rather
Coulomb fields for $Z^0$, $W^{\pm}$ and the photon to be rather
small, in as far as these exchanges are proportional to the fine
structure constants for the $SU(2)$ and $U(1)$ gauge groups in the
standard model, which are rather small. We should bear in mind
that we already {\em have} included the scalar components of these
{\em a priori} weak interaction gauge bosons. They were, namely, the
so-called eaten Higgs exchanges, which were supposed to be larger
because, as is explained in Appendixes D and E, the exchange of a
scalar component becomes independent of the fine structure
constants and is only given by the Yukawa coupling of the top
quark.

So here we want to discuss, as a small correction, the inclusion
of the timelike component exchanges of the weak gauge bosons. The
exchange of a $W$ boson has a similarity with the exchange
of the eaten Higgs in that it converts a top quark into a bottom
quark or oppositely. We could therefore roughly imagine that a
timelike $W$ exchange could---ignoring for the present what are
left and what are right components of the quarks---take the place
of an eaten charged Higgs component.

Very crudely we might therefore first simply imagine to include
the $W$ and $Z^0$ exchanges, by enhancing appropriately the eaten
Higgs couplings analogous to the gauge particle time components in
question. In the usual language, this means approximating the
exchange due to the timelike components of the gauge particles by
correcting by an overall factor the exchange due to the
scalar components alone (which is the one we call the eaten Higgs
exchange). Now the effective Coulomb potential for the eaten Higgs
is \be -\frac{g_t^2/2}{4\pi r}. \ee In the same notation the
Coulomb potential corresponding to the exchange of $W$'s 
between---now only left-handed---quarks becomes \be -\frac{g_2\tau^a/2 *
g_2\tau^a/2}{4\pi r} \ \mbox{(only for left-handed quarks)}, \ee
where, experimentally, we have at the $m_{Z}$ scale \be 1/\alpha_2
=\frac{ 4\pi}{g_2^2} \approx 30. \ee Here we have crudely included
$W^0$ exchange.

For a crude estimate we may take it that, in the roughly
nonrelativistic situation, there should be equally many
left-handed and right-handed top quarks, so that we can say in
the eaten Higgs exchange case, we have to start a box loop with
external right-handed top components. In analogy we have with the
time components to start with left-handed top components, but that
has approximately the same probability.

Using this way of arguing, we can effectively replace the $\tau^a$
matrices by the number $1$. We then get that the ratio of the
time-component exchange potential to the eaten Higgs exchange
potential is given by the factor \be \frac{1/30 *
(1/2)^2}{(g_t^2/2)/(4\pi)} \approx \frac{1/120}{1/28} = 0.237
\label{timetoeat}\ee Thus we may take a box loop, as discussed in
Sec.~4, to have its two eaten Higgs propagators increased by
the correction factor $1 + 0.237$. This would mean that, provided
we had external top-quark states being guaranteed to be a certain
linear combination of left-handed and right-handed components
corresponding to that for nonrelativistic particles, the box
diagram would be increased by a factor $(1+0.237)^2$. Hence the
critical $g_t$, namely $g_t|_{phase \ transition}$, should be
decreased by the fourth root of $(1+0.237)^2$, meaning
percentwise a {\em decrease} by $\ln(1+0.237)/2 = 12 \%$.

This is though an overestimate of the effect, because of the
following ``troubles'':

1) Our estimate of getting the squared correction factor
$(1+0.237)^2$ presupposes that interference terms, in the sense of
box diagrams with one time component and one eaten Higgs in the
same diagram, are really present.

2) There are not quite four $W$'s corresponding to the, in total,
four Higgses to be exchanged (as eaten Higgses or the original
Higgs).

If we have to give up the interference term, the factor
$(1+0.237)^2$ must be replaced by $ 1 + 0.237^2$ = $1.056$,
meaning only correcting the critical $g_t|_{phase \ transition}$
by {\em decreasing} it by $5.6 \%/4$ =  $1.4 \%$. The fact that we
have only 3 W bosons rather than 4 means that we should reduce
the ratio factor 0.237 from Eq.~(\ref{timetoeat}) to 3/4 of this
number. That would alone bring the above $12 \%$ down to $3/4
* 12 \% = 9\%$

So we take the correction coming from the exchange of the timelike
components of the SU(2) gauge bosons to be between $9\%$ and $3/4
* 1.4\% = 1.1 \%$. In other words we take the correction to give
a decrease of $g_t|_{phase \ transition}$ by $5\% \pm 4\%$.

In this crude estimate we really included the exchange of $W^0$, which
corresponds to a superposition of $Z^0$ and the photon $\gamma$.
Thus we are still left with having to include the orthogonal
$U(1)$ superposition of $Z^0$ and $\gamma$ in the next section.

\section{ U(1)-gauge boson exchange} \label{s11}

The photon or better the $U(1)$-gauge boson exchange
(a certain superposition of the photon and the $Z^0$
though mainly being the photon) may best be treated as
effectively modifying the gluon coupling, since it couples
similarly to the gluon.

The effective fine structure constant for the gluons, including
the 4/3 from Eq.~(\ref{ett}), is 0.109 *4/3 = 0.145 = 1/6.88,
which is to be compared with the inverse fine structure constant
for the $U(1)$ gauge group in the standard model $1/\alpha_1
\approx 100$ in the $Z^0$ mass region. This means that the
potential from the $U(1)$-gauge boson exchange is down by a factor
of 14.5 compared to that from the gluons. Since we found that the
gluons make up about one-third of the potential for binding, we
need only half of the 1/14.5 change in the $g_t^2$. In other words
we must correct $g_t|_{phase \ transition}$ by a relative change
of 1/2* 1/2 *  1/14.5 = 1/58. This means that the correction
coming from the inclusion of the $U(1)$ gauge particle exchange
causes the predicted critical $g_t$ to be {\em decreased} by $1/58
= 1.72 \%$

\section{Renormalization Group scale discussion}\label{s12}

The top-quark Yukawa coupling is strictly speaking a
running coupling constant, and we should use its running value at
the scale given by the typical momentum transferred by the
Higgses, which are emitted in the scattering processes relevant
inside the bound state. We have already found this to be $m_t$ in
Eq.~(\ref{e48}). This typical momentum transfer is also
crudely given by the inverse radius of the bound state which is,
as already estimated in Appendix C, of the order of $(\sqrt{4/3}
m_t)^{-1}$. That is to say that the critical Yukawa coupling
$g_t|_{phase \ transition}$, which we estimate above, is to be
interpreted as the running coupling at just the typical momentum,
or by the radius given scale, $\mu$ $\approx m_t $.

Usually the experimental result for the top-quark Yukawa coupling
$g_t$ is quoted as a running coupling at a
 scale of $\mu = m_t$, by making corrections to the measured
 pole mass. This gives the ``experimental''
 value $g_t(\mu = m_t) = 0.935$, whereas a more naive extraction
 from the measured mass \cite{mtop} of $172.6$ GeV gives $g_t|_{naive} = 0.992$.
 The formula used to get the running mass $m_t$ from the ``naive'' pole mass
 $M_t = g_t|_{naive} <\phi_h>/\sqrt{2}$ is \cite{runpole}
 \be
 m_t(M_t) = M_t \left [ 1 -1.333 \frac{\alpha_s(M_t)}{\pi}
 -9.125\left ( \frac{\alpha_s(M_t)}{\pi}\right ) ^2 \right ].\label{polecor}
 \ee
The effect described in this formula is that the top quark found
experimentally is a ``bare'' top quark surrounded by some gluons.

Now the question is to what extent the top quarks in the bound
state are also surrounded by gluons in the same way. Because from
outside, at distances large compared to its radius, the total
bound state is seen as a colorless particle, there must be such a
destructive interference between the gluons from the different
quarks or antiquarks that there will be no gluons at distances
much bigger than the radius. But that means that there are to
first approximation no gluons surrounding the quarks, when they
are inside the bound state. Thus the bound quarks are, from the
viewpoint of Eq.~(\ref{polecor}) the ``bare'' ones, described
by the running mass. This is the reason that we shall, in first
approximation, compare our prediction to $g_t(\mu = m_t) = 0.935$
rather than to the naive value $g_t|_{naive} = 0.992$.

By accident the scales associated with our critical coupling
$g_t|_{phase \ transition}$ and the experimental running mass are
essentially the same. So we do not need to make any
renormalization group correction. Nonetheless there is an
ambiguity in defining the precise scale and we will take a typical
uncertainty in the definition to be a factor of square root of 2.
In order to calculate the change $\Delta g_t$ in the top-quark
coupling generated by a shift in the scale $\mu$ by a factor
$\sqrt{2}$, we need to use the  $\beta$-function \be \frac{dg_t}{d
\ln \mu} = \frac{g_t}{16\pi^2} \left ( \frac{9}{2} g_t^2 - 8 g_3^2
- \frac{9}{4} g_2^2 - \frac{17}{12} g_1^2\right ). \ee
 Here $g_3=g_s$, $g_2$ and $g_1$ are the $SU(3) \times SU(2) \times U(1)$ running
 gauge coupling constants, related to their associated fine structure constants
 by $\alpha_i = g_i^2/(4\pi)$.

 Using the experimental values $g_t =
 0.935$, $g_3^2 = 4\pi *0.109$, $g_2^2  =  4\pi /30$ and
 $g_1^2  =  4\pi/100$ and taking $\Delta \ln{\mu} = \ln(2)/2 =
 0.347$, we get
that $\Delta \ln g_t = 0.347 * ( 3.9340 -   10.958 -   0.942
-0.178)/(16\pi^2)  = 1.79 \%$. Rounding this off to $2\%$, we
claim that the result for the correction due to the
renormalization group running scale is just $0\% \pm 2\%$.

\section{Collecting results.}\label{s13}

\begin{table}[ht]
  \centering
  \begin{tabular}{|c|c|c|c|}
\hline
 Name of correction     & Section    &Logarithmic \%     &
Estimated theoretical uncertainty in \%  \\
\hline
 Adjustment eaten      &    \ref{s4}  &   $4.0$    &  $8.0$  \\
\hline
  Higgs mass  &     \ref{sec5}   &   $5.2$    &  $4.0$  \\
\hline
 $s$-channel   &     \ref{s6}  &   $-18.8$    &  $6.0$  \\
\hline
 $m_t$ field dependence  &     \ref{s7}   &   $2.0$    &  $3.0$  \\
\hline
 Finite speed   &     \ref{finitespeed}   & $-17.2$ &  $6.0$ \\
\hline
 Many body &     \ref{s9}   & $14.2$  &  $4.6$ \\
 \hline
 $Z^0$ and $W$ exchange & \ref{s10} & $-5.0$ & $4.0$\\
 \hline
 $U(1)$-gauge exchange & \ref{s11} & $-1.7$ & $0.6$\\
 \hline
 Renorm group & \ref{s12} & $0.0$ & $2.0$\\
 \hline
 Total & & $-17.3$ & $14.2$ \\
\hline
\end{tabular}
\caption{Collecting corrections. } \label{table}
\end{table}

It is the value $g_t|_{phase \ transition} = 1.19$ from
Eq.~(\ref{bcorrected}) that has to be changed by the total
correction factor, resulting from all the corrections to
eq.~(\ref{bcorrected}) discussed throughout the paper and
presented in Table 1. The collected ``total'' percentwise
logarithmic correction turns out to be $-17.3 \% \pm 14.2 \%$.
Thus the running Yukawa coupling for the top quark at the $m_t$
scale is predicted, under our basic assumption that the mass of
the bound state shall be just tuned in---mysteriously---to be
small, to be \be g_t = g_t|_{phase \ transition} = \exp(-17.3 \%)
* 1.19 = 1.001 \pm 14.2 \% = 1.00 \pm 0.14 \ee This result is to
be compared with the value from experiment \cite{mtop}, obtained
from a top-quark pole mass of $172.6 \pm 1.4 $ GeV: \be g_t(m_t) =
0.935 \pm 0.008. \ee This means that our prediction from the
masslessness of the bound state is fulfilled up to $ (1.001 -
0.935)/0.142$ = $0.46$ standard deviations.

At least this calculation means that the very exotic bound state
we propose has a mass squared which is down by a factor of the
order of $0.142$ relative to the natural mass squared scale for
this type of bound state, namely, the mass squared of $12m_t$,
i.e.~$144 m_t^2 \approx 4$ TeV$^2$. That is to say that  the mass of
the bound state must be at least as small as of the order $ 12 m_t
* \sqrt{4*0.142} = 1560$ GeV. However, the real point is that it
could easily within the errors be much lighter, e.g., zero mass.
Looked upon as such an estimate of the bound state mass, our
calculation at first sight appears to not be so impressive.
However, it means that if the top-quark Yukawa coupling $g_t$
deviated outside our error estimate, it would be likely that
either (i) there would be such strong binding that a condensate
would unavoidably have formed and we would live in a phase with
such a condensate safely dominating, or (ii) the binding would be
very tiny compared to $12m_t$. Essentially the case of binding
with a binding energy just of the order of 12$m_t$ is what
corresponds, in the $g_t$ formulation, to a rather narrow and
impressive range in which the experimental coupling quite
remarkably lies.

The really remarkable thing coming out of our calculation is that,
by requiring the masslessness for the bound state, we get the
empirical Yukawa coupling with such a high accuracy of $14.2 \%$
that there is a rather striking agreement. In itself it is
remarkable even to get the right order of magnitude for the Yukawa
coupling. However the fact that we get it just right with a
$14.2\%$ accuracy is something that would only occur, even if the
agreement within a factor $e$ were already guaranteed, in one
out of 8 cases. So it almost calls for some underlying theory to
explain that coincidence. We would say that the
``multiple point principle'' of requiring many vacua with the same
energy density \cite{itepportoroz,brioni} would function as such a
theory, if we, namely, take there to be two vacua, one with and one
without a bose condensate of the bound state discussed in this
article. We can say it is a case of a strange fine-tuning of the
Yukawa coupling for the top quark and that a fine-tuning machinery
is called for.

\section{Conclusions}

The main content of the present article has been a calculation,
performed inside the standard model, of the mass of a special
bound state. In fact we calculated the mass of the bound state
formed from 6 top quarks and 6 antitop quarks. The importance of
just this set of quarks and antiquarks is that they form a closed
shell, so that there is a significant decrease in the strength of
binding when the next quark or antiquark is added to the system.
The remarkable result we found is that the top-quark Yukawa
coupling experimentally has just the value that allows this bound
state of the $6t + 6 \overline{t}$ to be totally {\em massless}.
That is to say, within the uncertainty, it can very easily be that
the binding is so strong as to just cancel the mass energy of the
constituents. In fact we formulated our calculation so as to
evaluate just that specific value of the top-quark Yukawa
coupling, which gives precisely zero mass for the bound state of
its 12 constituents. It must be admitted that this conclusion
of the 6 top and 6 anti-top quark state even binding---let alone
so strongly as to get zero mass---is at variance with the conclusion
of Kuchiev, Flambaum and Shuryak \cite{Shuryak} who do not even have it
bind. But we have included several further important effects---such as
eaten Higgses and corrections to the Higgs mass to be used {\em inside}
the bound state---in our calculation
of the binding strength. As suggested by a toy model calculation in
Appendix J, there is reason to believe that the mass of the bound 
state---including the question of binding---has a kink behavior as a function
of, e.g., the Yukawa coupling $g_t$. So two calculations performed
on different sides of such a kink value of the variable $g_t$
could {\em a priori} give quite different results. Depending on where
exactly ``the phase transition'' is only one of two such calculations
would be correct, the other one being analogous to calculating
the properties of fluid superheated water for a temperature where the
true phase is the vapor phase.

So we can suspect that, provided one included enough of our above-mentioned
corrections, the correct ``phase'' for
the calculation would be the one with a ``collapsed'' Higgs field
inside the hoped for bound state. However Kuchiev et al.~\cite{Shuryak}
made their calculation in the phase with an uncollapsed Higgs field,
i.e., the calculation is in the wrong phase. But presumably, without the
inclusion of eaten Higgs exchange and our other corrections,
the experimental value of $g_t$ would lie in the phase in which
Kuchiev et al.~worked.

Our aim was then to see to what accuracy the rather mysterious
coincidence of bound state masslessness actually works
in the phase with a collapsed Higgs field. So we
wanted to compute the critical Yukawa coupling $g_t|_{phase \
transition}$, defined here as the one making the bound state
massless, as accurately as possible. We did that by first
calculating it in a rather crude way, leading to the value
$g_t|_{phase \ transition} = 1.19$ in Eq.~(\ref{bcorrected}).
In this first step in the calculation we included both Higgs
exchange and gluon exchange in the $t$ channel and the $u$ channel,
but did not yet include the $s$-channel exchange (which is more
difficult to calculate); we also very crudely corrected for the
fact that there could also be left-handed $b$ quarks and
antiquarks virtually present in the bound state, essentially
replacing the $t$ quarks from time to time. We also, in this first
calculation, used a very crude approximation of letting each quark
encircle a conglomerate of all the other 11 quarks concentrated
into one point. However we did take the double counting into
account and thus really calculated with only 11/2 particles in the
center. We made the calculation totally nonrelativistically, as
is almost needed to calculate a bound state without having to
truly go to the Bethe-Salpeter equation.

After this first calculation, we then made a series of 9
corrections listed in Table 1 in the foregoing section. Together
these corrections led to lowering the predicted critical Yukawa
coupling by $17.3 \%$ counted  logarithmically. The resulting
Yukawa coupling that would give just zero mass for the bound state
of $6t + 6\overline{t}$ was thus computed to be $ g_t|_{phase \
transition} = 1.00 \pm 0.14 $. This uncertainty of $14\%$ is only
a very crude estimate of the uncertainties in the many corrections
added up in quadrature. At first sight this $14 \%$ uncertainty
appears to be a small error. However, one should bear in mind
that what we really estimate is $g_t^4$ rather than $g_t$ itself.
Therefore the true uncertainty on our calculation, namely, for $g_t^4$,
is in fact rather of the order of $70 \%$.
We performed the calculation so as to
estimate the running Yukawa coupling at the $m_t$ mass scale,
where the pole mass correction performed on the experimentally
measured top-quark mass leads to the experimental running Yukawa
coupling value $g_t(\mu = m_t) = 0.935 \pm 0.008$. This
experimental value has thus fallen, within an
uncertainty of only $0.46$ standard deviations, on the value needed
to make the bound state massless.

\section*{Acknowledgements}

We would like to thank Roman Nevzorov for many helpful discussions.
CDF thanks the Niels Bohr Institute and HBN thanks CERN for hospitality
while working on parts of this paper.

\section*{ Appendix A: Notation}
\label{A1} In Sec.~\ref{sec2} we work with just one Hermitian
(``real'') field for the physical Higgs particle $\phi_h$. Here we
shall give some notation for this field and the related complex
doublet field $\phi_H$.

We take the Lagrangian for the real field $\phi_h$ to be
normalized as \be {\cal L}(x) =
\frac{1}{2}(\partial_{\mu}\phi_h)^2 +
\frac{1}{2}|m_{hb}|^2\phi_h^2 -\frac{\lambda}{8}\phi_h^4
+\frac{g_t}{\sqrt{2}} \overline{\psi}_t \psi_t \phi_h +
\overline{\psi}_t\gamma^{\mu}\partial_{\mu}\psi_t+... \label{yh}
\ee Phenomenologically we know that the vacuum expectation value
of the Higgs field is \be <\phi_h> = |m_{hb}|/\sqrt{\lambda/2} = v
= 246\ \mbox{GeV}, \ee while the physical Higgs mass becomes \be
m_h = \sqrt{2}|m_{hb}| = \sqrt{\lambda} \, v. \ee So, for example,
for a Higgs mass of $m_h = 115$ GeV, we find in this notation that
$ \lambda = \frac{(115 \ GeV)^2}{(246 \ GeV)^2} = 0.218$

In order to treat the eaten Higgses too, as we do in Sec.~\ref{sec3}, 
we must introduce the Higgs doublet complex field notation
in which \be \phi_H = \left ( \begin{array}{c}\phi^+ \\
\phi^0\end{array} \right ), \ee where then we take \be \phi^0 =
\frac{1}{\sqrt{2}}(\phi_h+ i \phi_2) \label{cr}, \ee with $\phi_h$
and $\phi_2$ real. With this relation we are then forced to take
the Lagrangian density for the complex field doublet to be \be
{\cal L_H }= |D_{\mu}\phi_H|^2 + |m_{hb}|^2 \phi_H^{\dagger}\phi_H
- \frac{\lambda}{2}(\phi_H^\dagger\phi_H)^2. \ee


With the substitution (\ref{cr}), the Yukawa interaction term in
(\ref{yh}) becomes \be {\cal L_H} = ...+g_t(\overline{\psi}_{tR}
\phi_H^{\dagger} \psi_{tbL} + h.c.)+... \label{yH} \ee Here we
have introduced the splitting of the Dirac spinor into its Weyl
representation components---meaning left and right being
considered separately---and also introduced the left-handed $b$
field, so that we have now a doublet of left-handed fields under
the weak isospin: \be \psi_{tbL} = \left (
\begin{array}{c}\psi_{tL}\\ \psi_{bL} \end{array}\right ). \ee We
also denote the right-handed components of the $t$ field by
$\psi_{tR}$.

\section*{Appendix B: Infinite momentum frame for nonrelativistic approximation and analyticity}

We shall here see how a nonrelativistic atomlike theory gets
written in the infinite momentum frame. Let us consider a cluster
of $n$ constituent particles numbered by $i = 1,...,n$ with masses
$m_i$ and longitudinal momenta $p_{zi}$ written as \be p_{zi} =
x_i p_z. \label{xpz} \ee
 Here $p_z$ is some very large momentum used to specify the very fast moving
 frame that is the IMF. Then the energy of the cluster
 of particles in this frame, in which
 we think and in which the particles move very fast, is expanded as follows
 \be
 E_{IMF \ cluster} = p_z + \left ( \sum_{i=1}^n \frac{m_i^2 + \vec{p}_{Ti}^2}{2x_i} \right )/p_z
 + \frac{1}{2}\sum_{i,j, \ i\neq j} V_{ij}/ \gamma_{ij}.
 \ee
We use the notation $\vec{p}_{Ti}$ for the transverse part of the
momentum of particle $i$ and $\vec{p}_T = \sum_{i=1}^n
\vec{p}_{Ti} = 0$. Here the nonrelativistic scalar potential
$V_{ij}$, for particle $i$ influencing particle $j$, is being
boosted from the cluster rest frame to the infinite momentum frame
and thereby Lorentz contracted. Because of the Lorentz contraction
of the wave function for particle $j$, its scalar interaction goes
down by the factor $1/\gamma_j$ = $\sqrt{1-v_{lj}^2}$, where
$v_{lj}$ is the longitudinal velocity of particle $j$. If we had
thought about the interaction the opposite way around, we would
have gotten $1/\gamma_i$ instead. But if the particles keep
interacting they must run with the same speed and that would mean
$\gamma_i \approx \gamma_j$, so that we can put $\gamma_{ij} $
equal to both of them.

Since the longitudinal $\gamma_i = p_z x_i/m_i$ in the infinite
momentum limit, we have in this case of the same longitudinal
velocity that $x_i/x_j = m_i/m_j$.

It is also obvious that (\ref{xpz}) implies the well-known normalization
\be
\sum_{i=1}^n x_i =1.
\ee

Especially in the first nonrelativistic approximation for the
internal motion of the cluster, the relative velocities are small
and thus the $x_i$'s are proportional to the corresponding
$m_i$'s. Also, in this first approximation of $x_i \propto m_i$,
we could for instance write, using the average of $1/\gamma_i$ and
$1/\gamma_j$ for $1/\gamma_{ij}$:
\begin{eqnarray}
 E_{IMF \ cluster} &= &p_z + \left ( \sum_{i=1}^n \frac{m_i^2 + \vec{p}_{Ti}^2}{2x_i}
 + \frac{1}{2}\sum_{i,j, \ i\neq j} V_{ij}*\frac{1}{2}( \frac{m_i}{x_i} +
 \frac{m_j}{x_j} )  \right )/ p_z\\
 &= &p_z + \frac{1}{2p_z}    \left ( \sum_{i=1}^n \left (\frac{m_i^2}{x_i} +
 2\frac{\vec{p}_{Ti}^2}{2x_i} \right )
 + \frac{1}{2}\sum_{i,j, \ i\neq j} V_{ij}*( \frac{m_i}{x_i} +
 \frac{m_j}{x_j} ) \right ).\label{EIMFa}
 \end{eqnarray}

Comparing with the IMF expansion (\ref{IMF}) of $E_{IMF}$, we see
that the squares of the eigenmasses for the bound states in the
channel considered are given as eigenvalues of the operator
$\left ( \sum_{i=1}^n \left (\frac{m_i^2}{x_i} +
 2\frac{\vec{p}_{Ti}^2}{2x_i} \right )
 + \frac{1}{2}\sum_{i,j, \ i\neq j} V_{ij}*( \frac{m_i}{x_i} + \frac{m_j}{x_j} ) \right )$,
 so that we determine the bound state masses from the eigenvalue
 equation:
 \be
 \left ( \sum_{i=1}^n \left (\frac{m_i^2}{x_i} +
 2\frac{\vec{p}_{Ti}^2}{2x_i} \right )
 + \frac{1}{2}\sum_{i,j, \ i\neq j} V_{ij}*( \frac{m_i}{x_i} + \frac{m_j}{x_j} ) \right )\Psi
 = m_{bound}^2\Psi.
 \label{eigen}\ee
 The remarkable thing for us here is that there is no obvious reason why this
 eigenvalue equation should have any singular behavior for $m_{bound}^2$ at zero.
 Therefore we expect that the eigenvalues, meaning the masses squared of the bound states, will behave smoothly
 as a function of the parameters such as $g_t$. That suggests confidence in using
 a low order Taylor expansion in the parameters, even when the bound state mass
 squared $m_{bound}^2$ comes close to zero.
 In other words we expect to have no singularities at
 $m_{bound}^2 = 0$, when we use the eigenvalue equation (\ref{eigen}) to obtain
 the IMF-mass squared of the bound state.

 In order to check that we do indeed get to the slightly surprising factor 1/2 in (\ref{factorhalf}),
 meaning that the binding energy in the formal nonrelativistic calculation should
 only compensate one-half of the mass in order to make the bound state just massless,
 we shall here take the nonrelativistic approximation to our IMF formalism:
 With  the nonrelativistic approximation in mind, in the frame of the bound state, we define
 the $\Delta x_i$'s by
 \be
 x_i =  \frac{m_i}{ \sum_{j} m_j}+\Delta x_i  \equiv x_{i \ old}+ \Delta x_i.\label{appx}
 \ee
 Below we shall prove that, by Taylor expanding the term $\frac{m_i^2}{x_i}$ in the expression
(\ref{EIMFa}) for $E_{IMF/cluster}$, we obtain the longitudinal part of the kinetic energy
 quite analogous to the transverse part already present.

Neglecting the $\Delta x_i$ and inserting $x_i = x_{i \ old}=
\frac{m_i}{\sum_j m_j}$ into (\ref{EIMFa}), we get
 \begin{eqnarray}
 E_{IMF \ cluster} & = &
 p_z + \frac{1}{2p_z}    \left ( (\sum_{i=1}^n m_i )^2 +
 2(\sum_k m_k) *\left (\sum_i\frac{\vec{p}_{Ti}^2}{2m_i}
 + \frac{1}{2}\sum_{i,j, \ i\neq j} V_{ij}\right) \right )\\
 & = & p_z + \frac{\sum_j m_j}{2p_z}\left ((\sum_k m_k) +
 2\left ( \sum_i \frac{\vec{p}_{Ti}^2}{2m_i}
 + \frac{1}{2}\sum_{i,j, \ i\neq j} V_{ij}\right) \right )\\ \label{eq123}
&=& p_z + \frac{1}{2p_z} \sum_{i=1}^n \frac{m_i}{x_{i \ old}}(m_i
+ 2H_i|_{\perp}), \label{eq124}
 \end{eqnarray}
where
\begin{equation}
 H_i|_{\perp} = \frac{\vec{p}_{Ti}^2}{2m_i}
 + \frac{1}{2}\sum_{j, \ i\neq j} V_{ij}.
\end{equation}

Actually we will now show that the Taylor expansion of the main
term $\left( \sum_i \frac{m_i^2}{x_i} \right)/(2p_z)$ in
(\ref{EIMFa}) has a dependence on the longitudinal momentum of the
constituents, which can be interpreted as the missing longitudinal
momentum dependent part of the kinetic energy, looking quite
analogous to the transverse part.

In the ``at rest" limit, in which the particles in the cluster lie
still in the cluster rest frame, we have $p_{zi}=x_i* p_z$ with $
x_i = x_{i \ old} = m_i/(\sum_j m_j)$. However, if the particles
are not at relative rest, the $x_i$'s will deviate from the $x_{i
\ old}$ as in (\ref{appx}): \be x_i =x_{i \ old} + \Delta x_i. \ee
Of course \be \Delta x_i = \frac{\Delta p_{zi}}{p_z}, \ee where
$\Delta p_{zi}$ stands for the deviation of the longitudinal
momentum of the $i$th particle from the value $ x_{i \ old} p_z =
\frac{m_ip_z}{\sum_j m_j}$, which is the momentum it would have in
the ``resting'' approximation. This $\Delta  p_{zi}$ could roughly
be considered to be the result of the boosting of the longitudinal
component of momentum $p_{nr\ i}$ of the particle $i$, measured in
the rest frame of the cluster, from the rest frame of the cluster
to the infinite momentum frame we consider. In the nonrelativistic
approximation, in which the velocity of this cluster rest frame is
given by the velocity $v$ with associated $\gamma$ satisfying
$v\gamma = p_z/( \sum_j m_j)$ (or approximately for very large
$p_z$ just $\gamma = p_z/(\sum_j m_j)$), we have \be \Delta p_{zi}
= \gamma p_{nr\ i} = \frac{p_z}{\sum_j m_j} p_{nr\ i}. \label{pzi}
\ee
 The Taylor expansion of the main $x_i$-dependent term in the $E_{IMF \ cluster}$ now gives
\begin{eqnarray}
\frac{1}{2p_z}\frac{m_i^2}{x_i} & = & \frac{1}{2p_z}\left (
\frac{m_i^2}{x_{i \ old}} - \frac{m_i^2}{x_{i \ old }^2} \Delta
x_i + \frac{m_i^2}{ x_{i \ old} ^3} \Delta x_i ^2 +...\right ).
\end{eqnarray}
With the insertion of (\ref{pzi}) into the third term in this expansion, which is
proportional to $\Delta x_i^2$, we get
\begin{eqnarray}
 \frac{1}{2p_z}  \frac{m_i^2}{ x_{i \ old} ^3} \Delta x_i ^2 &  = &
\frac{1}{2p_z} \frac{m_i^2}{ x_{i \ old} ^3} \frac{\Delta p_{zi}^2}
{p_z^2}\\ & = &  \frac{1}{2p_z^3} \frac{(\sum_j m_j )^3}{m_i} \Delta p_{zi}^2 \\
& \approx & \frac{p_{nr\ i}^2}{2p_z x_{i \ old}}.
\label{longitudinal}
\end{eqnarray}
We see that this term (\ref{longitudinal}) is precisely analogous
to the transverse term  $ \frac{p_{Ti}^2}{2p_z x_i}$. The second
term in the Taylor expansion, the one going linearly in $\Delta
x_i$, is quickly seen to be proportional to the sum $ \sum_i
\Delta x_i $ which is zero, provided one keeps to the
normalization $ \sum_i x_i = 1$.

Now, adding these terms proportional to $\Delta x_i^2$ to
(\ref{eq124}), we obtain
\begin{equation}
E_{IMF \ cluster} = p_z + \frac{1}{2p_z} \sum_{i=1}^n
\frac{m_i}{x_{i \ old}}(m_i + 2H_i), \label{eq128}
\end{equation}
where
\begin{equation}
 H_i = \frac{\vec{p}_i^2}{2m_i}
 + \frac{1}{2}\sum_{j, \ i\neq j} V_{ij}.
\end{equation}
Here $\vec{p}_i^2 = \vec{p}_{Ti}^2 + p_{nr\ i}^2$ is the total
momentum squared of particle $i$ in the cluster rest frame.

\section*{Appendix C: Radius estimate}

In order to estimate the radius of our bound state in the critical
coupling case, we may use Eq.~(\ref{e18}) and the virial theorem.
From the virial theorem for a $1/r$ potential, it follows that the
total binding energy comes about by the average of the potential
energy making up twice the binding energy (being negative like the
binding energy), while the kinetic energy is numerically equal to
the binding energy but is positive and thus compensates away
one-half of the potential energy. Now, according to (\ref{e18}),
the binding energy per constituent particle must be $m_t/2$ in the
critical case. It therefore follows, from the above virial theorem
consideration, that we must have \be m_t/2 = <T> =
<\frac{\vec{p}^2}{2m_t}> = \frac{<p_x^2> + <p_y^2> +
<p_z^2>}{2m_t}. \ee For symmetry reasons it then follows that \be
<p_x^2> = <p_y^2> =<p_z^2> = \frac{m_t^2}{3}.\label{px2} \ee

Now we want to use the fact that, in the ground state, the
Heisenberg uncertainty relation \be <x^2> <p_x^2> \, \ge \, 1/4
\ee is actually an approximate equality, so that we really have
\be <x^2> <p_x^2> \approx 1/4. \label{Heisenberg} \ee The ground
state of a system like ours, or an atom, is achieved by
concentrating the constituents with minimal energy as closely
together as the Heisenberg uncertainty relation allows. Now the
true equality is achieved only for a Gaussian wave function.
However the deviation from Gaussian form only comes in to second
order (in some parameter measuring the deviation from the Gaussian
form of the wave function) because, imagining an abstract Taylor
expansion for the deviation from the Heisenberg uncertainty
relation, it could only have second order terms without violating
the inequality.

Inserting $<p_x^2>= m_t^2/3$ from (\ref{px2}) into
(\ref{Heisenberg}), we get in the ground state in the critical
case \be <x^2> = <y^2> = <z^2> \approx \frac{3}{4m_t^2}. \ee From
here \be <r^2> =3<x^2>\approx \frac{9}{4m_t^2}.\label{r2} \ee With
the wave function $\psi \propto \exp(-r/r_0) $, one easily finds
\be <r^2> = 3 r_0^2 \label{r2r0} \ee and so derives that \be r_0
\approx \sqrt{3/4} \frac{1}{m_t} \ee from (\ref{r2}) and
(\ref{r2r0}).

Note that this argument is true independent of whether we have
gluon or Higgs exchange or a mixture, provided the exchanged
particle is sufficiently light so that the scaling properties
assumed about the potential, when using the virial theorem, remain
valid.

\section*{Appendix D: Eaten Higgses from $W$ and $Z�$}\label{appendix3}

We shall explain how the nonconserved part of the current,
coupling to $W$, causes a propagator to be inversely proportional
to the gauge coupling squared  for small four-momentum transfer
$q^2$ and which, thus, can cancel the squared gauge coupling
coming from the vertices. In this way we can, even in the limit of
the gauge coupling going to zero, have a nonzero exchange force
due to the gauge particles $W$ and $Z^0$.

When, for instance, a top quark is converted into a bottom quark
by emission of a $W$, the transition current $ j^{\mu}_{W^{+}} =
\frac{g_2}{\sqrt{2}}\overline{\psi}_{bL} \gamma^{\mu} \psi_{tL}$
is not conserved due to the masses of the top and bottom quarks.
[Here we took the general $W$ current to be $j^{a \, \mu} =
\frac{g_2}{2} \overline{\psi}_{btL}\gamma^{\mu} \tau^a \psi_{btL}$
and the normalization $W^+$ = $ \frac{1}{\sqrt{2}}(W_1 +i W_2)$,
so that $W^+$ will couple to the current $j_{W^+}^{\mu} =
\frac{g_2}{\sqrt{2}} \overline{\psi}_{bL}\gamma^{\mu}\psi_{tL}$.]
In fact, using the equations of motion for the quarks in the
background of the Higgs field, the divergence of the current
becomes \be
\partial_{\mu}j^{\mu }_{W^+}=\partial_{\mu} j^{\mu}
= -i\frac{g_2}{\sqrt{2}}m_t \overline{\psi}_{bL} \psi_{tR}
 +i\frac{g_2}{\sqrt{2}}m_b \overline{\psi}_{bR}\psi_{tL}\neq 0.
\ee Here $m_t$ and $m_b$ are the top and bottom quark masses,
which are given by \be m_t = \frac{g_t}{\sqrt{2}} <\phi_h>;\quad
m_b = \frac{g_b}{\sqrt{2}} <\phi_h>. \ee (If we consider the $b$
quark massless as a good approximation, then $g_b = 0$ and
$m_b=0$.)

Considering the inverse propagator for the gauge boson, as
obtained from $ -1/4 * F^i_{\mu\nu} F^{i\, \mu\nu}
-m_W^2A^i_{\mu}A^{i\, \mu}$, we see that the kinetic part of this
inverse propagator can be zero for currents having the direction
of a four gradient (as is a consequence of the gauge invariance of
this kinetic part). Thus the propagator goes as the inverse of
$m_W^2$ in such cases. But now the $W$ only got its mass $m_W$
nonzero due to the Higgs field and this mass is actually
proportional to the gauge coupling $ m_W = \frac{g_2}{2} <\phi_h>
= \frac{g_2}{\sqrt{2}} <\phi_H>$. Thus the propagator for
(nonconserved) currents not coupling to the kinetic part of the
inverse propagator becomes proportional to the inverse gauge
coupling constant squared $\propto 1/g_2^2$. The exchange
amplitude for this nonconserved contribution thus has its $g_2$
dependence canceled. This then means that, even in the limit of
the gauge coupling $g_2 \rightarrow 0$, the exchange of the
massive gauge boson cannot be ignored when the current is not
conserved. In this limit the exchange amplitude can only depend on
the other coupling constant, the Yukawa coupling, and indeed it is
physically really just the exchange of the eaten Higgs components
that comes out of this limit from the gauge particle exchange.

The conclusion we want to draw in this Appendix is this: In the
formalism in which one considers massive gauge bosons, such as $W$
and $Z^0$, one only has to consider, in addition, the {\em
physical} Higgs particle components. However, in the limit of
letting the gauge couplings $g_2$ and $g_1$ go to zero, the gauge
boson exchange {\em does not fully decouple}. Rather, in this
limit, the gauge boson exchange simply becomes what one would get,
in addition to the physical $\phi_h$ Higgs exchange contribution,
by including the full Higgs field $\phi_H$ with all its four real
components. So in this limit one is truly led to the pure Higgs
model, {\em but with all the components, including the previously
eaten ones}.

   Although the above argument was very suggestive of what likely goes
   on in the limit of very weak gauge couplings, we actually should check that
   we do obtain the correct exchange amplitude corresponding to the eaten
   Higgs. A simple check of this can be done as follows, if it is
   accepted that we can be allowed to talk about the nonconserved part of the
   current and take it to be in momentum representation:
   \begin{eqnarray}
    j^{\mu}_{W^+}|_{nonconserved} =   j^{\mu}|_{nonconserved} &= &\frac{1}{q^2}q^{\mu} q_{\nu} j^{\nu} \\
   = \frac{1}{q^2} q^{\mu}\frac{1}{i} \partial_{\nu} j^{\nu}
   &= &\frac{1}{q^2} q^{\mu}
(-\frac{g_2}{\sqrt{2}}m_t \overline{\psi}_{bL} \psi_{tR}
+\frac{g_2}{\sqrt{2}}m_b \overline{\psi}_{bR}\psi_{tL}).
   \end{eqnarray}
Using this ``only nonconserved part of the current'' together with
a W propagator put equal to just $\frac{1}{m_W^2}$, as is expected
to be a good approximation for $g_2 \rightarrow 0$, we can
formally obtain an expression for the amplitude corresponding to
the W-exchange diagram for the scattering of a pair of quarks. For
the quark transitions from $t$ to $b$ and oppositely from $b$ to
$t$ at the two vertices and putting $m_b =0$ for simplicity, we
get the following expression:
\begin{eqnarray}
\frac{j^{\mu}_{nonconserved} j_{\mu \; nonconserved}^{\dagger} }{m_W^2}& \approx & (g_2/\sqrt{2})^2 m_t \overline{\psi}_{bL} \psi_{tR}
* m_t  \overline{\psi}_{tR} \psi_{bL} * \frac{1}{q^2 m_W^2} \\
&\approx & {g_t} \overline{\psi}_{bL} \psi_{tR} * \frac{1}{q^2}{g_t} \overline{\psi}_{tR} \psi_{bL}.
\end{eqnarray}
This scattering amplitude for the quark transitions is precisely
what you get by exchange of an ``eaten'' Higgs.

\section*{Appendix E: Counting eaten Higgses}\label{appendix2}

In Sec.~\ref{sec3} we introduced the extra Higgs components, which
are eaten by the gauge particles, together with the $b$ quark.
They were used to consider (formally) the contribution of a box
diagram to the elastic scattering of the weak singlet right-handed
top quark $t_R$ and its antiparticle $\overline{t_R}$. Because of
the fact that there were now 4 real components of the complex
doublet Higgs field $\phi_H$ propagating in the loop rather than
just the one real physical Higgs field $\phi_h$ considered in 
Sec.~\ref{sec2}, we argued that the scattering amplitude must increase
by a factor of 4. In this Appendix we shall now confirm this
factor of 4, by simply evaluating the ratio of box-diagram
amplitudes for $t_R \overline{t_R}$ scattering in the two cases:
(i) including only the physical Higgs component $\phi_h$ and the
left-handed top quark $t_L$ in the loop, and (ii) including all 4
components of the Higgs field $\phi_H$ and both the left-handed
top $t_L$ and bottom $b_L$ quarks in the loop.\footnote{Note that
when we formally only consider the left-handed quarks in a loop,
it means that we have ignored the quark mass and left it as a
perturbation to be considered later.}

Using the notation of Appendix A, we see that the box
diagram---with two left-handed top or bottom quarks and two
Higgses in the four-sided loop---gets changed in the following
ways, when going from case (i) including only the physical Higgs
component to case (ii) including the full 4 component Higgs field
of the standard model:

1) We get rid of the $1/\sqrt{2}$ in the Yukawa coupling Lagrangian density.
This means that the scattering amplitude goes up by a factor of $\sqrt{2}^4$ relative
to case (i) with only the physical Higgs.

2) After including all four components, we have to evaluate an
SU(2) trace for the box diagram, corresponding to the fact that a
weak isodoublet circles around the box loop. This means that the
amplitude goes up by a factor of 2.

3) There is a type of diagram which is allowed for case (i) with
the physical Higgs field $\phi_h$ alone, but which is forbidden
for case (ii) when we consider the complex doublet Higgs field
$\phi_H$ which carries a charge of weak hypercharge. In fact there
is for case (i), with only the physical Higgs field being
considered, the possibility of ``crossing'' the two Higgs
propagators in the box diagram. Because of this possibility, we
get a factor of 2 bigger amplitude for the case (i). This means
that going from case (i) to the four component case (ii), one gets
a factor of $1/2$.

Altogether we thus get an increase by a factor of $\sqrt{2}^4 * 2
* \frac{1}{2} = 4$, by including all four components instead
of just the physical Higgs, and that is just what we argued for in
Sec.~\ref{sec3}.

\section*{Appendix F: Distribution of lengths of loops}

In Sec. \ref{s4} we made the assumption that, without the
weighting coming from the number of isodoublet states that can
circle in a loop of $n$ ``propagators'', the number of such loops
statistically had a smooth distribution as a function of $n$,
although there only are loops with an even number $n$.

We here want to consider this assumption in a little more detail:
Imagine that we construct a random diagram by going along in small
steps following the construction of a loop of propagators for the
isodoublet particles [i.e., left-handed $b$ or $t$ quarks or Higgs
particles (including the eaten Higgses)]. Then as one goes along
it is sensible to think that, almost all the time, there is the
same chance of getting back to the starting point of the initiated
loop. This chance of getting back to the starting point should,
namely, all the time be roughly 1 divided by the number of
possible attachment points (say the order of the diagram) for an
isodoublet propagator getting inserted. We must admit however that
we have not clearly stated which way one should imagine to build
up the diagram. One way would be to imagine that the structure of
the diagram is already given and one just successively attaches a
label, doublet or singlet, to the propagators in an already given
diagram. One would still have to think of the given diagram
statistically only and that the chance for the doublet loop being
followed reaching any {\em a priori} vertex could be taken to be
the same all through the construction. Then, although in principle
the possible attachment points at any stage of the construction
become all the vertices not yet used, we do not correct for this
fact that vertices already used are no longer accessible.

This crude argumentation will give an exponentially decaying
distribution for the distribution of the loop length $n$. However
it will fall off so slowly with large $n$ that the average loop
length gets of the order of the full number of attachment
possibilities. This means, assuming a large diagram, a very flat
distribution for the first few $n$ values to the extent that they
can at all be realized. (For instance, $n=2$ would only occur
inside self-energy diagrams for the right-handed top quark, and
also only even $n$ are possible.)

\section*{Appendix G: Flattening of potential for small $r$}
\label{app5}

For the purpose of estimating an effective Higgs mass to take into
account the difference between the Yukawa and Coulomb potentials,
we want first to estimate how the Higgs field varies with the
distance $r$ from the center of the bound state. Strictly speaking
we should calculate the wave function distribution for the
constituents and evaluate the Higgs field with this density of
constituents used as the source. However, we shall here
approximate the correct density distribution by a distribution,
$\rho_0$, that is constant in 3-space inside a radius $R$, i.e.,
for $r<R$, and zero outside. Its value is chosen so as to
correspond to there being a ``charge'' (i.e.~the number of
constituents times $g_t/\sqrt{2}$) of $(11/2) g_t/\sqrt{2}$. We
shall take the $R$ parameter then to be the average radius of the
wave function distribution, i.e., \be R = <r> = \frac{3}{2} r_0.
\ee

In the very center there must (statistically) be a certain density
of constituent particles having an extremum there. This means
that, in the immediate neighborhood of the center, the density of
constituents goes as \be \rho(r) \approx \rho(0). \ee This leads
to a spherically symmetric potential or Higgs field, satisfying
the Laplace equation with the source term \be g_t/\sqrt{2}
* \rho = \rho_0 = \frac{1}{r^2} \frac{\partial }{\partial r} \left
( r^2 \frac{\phi_h}{\partial r}\right ). \ee The resulting
potential is \be V = \frac{\rho_0}{6} r^2 +C \quad \hbox{for} \ r
\le R. \ee The physical number of constituents inside the average
radius $R$ is half the number of constituents in total and we thus
identify it with our number of effective constituents in the
center $Z=11/2$. Thus we find \be \rho_0 \approx
\frac{Zg_t/\sqrt{2}}{4\pi R^3/3} \approx \frac{11/2 *
g_t/\sqrt{2}}{4\pi/3 * <r>^3}.\label{rho0} \ee If we also use
Z=11/2 for the outside field, being approximated as a Coulomb
potential, we shall automatically get that the slope of the
potential is continuous: \be
V = \left \{ \begin{array}{rcl}\frac{\rho_0}{6} r^2 +C & \hbox{for}& r \le R\\
-\frac{Zg_t/\sqrt{2}}{4\pi r}& \hbox{for} & r \ge R\end{array} \right .
\ee
Inserting (\ref{rho0}) and adjusting $C$ to make the two expressions coincide
for $r=R$ leads to

\begin{eqnarray}
V& =& \left \{ \begin{array}{rcl}\frac{Zg_t/\sqrt{2}}{8\pi R^3}( r^2 - 3R^2)
& \hbox{for}& r \le R\\ -\frac{Zg_t/\sqrt{2}}{4\pi r}& \hbox{for} & r \ge R\end{array} \right .\\
 &= & \frac{Zg_t/\sqrt{2}}{8\pi R^3}\left \{ \begin{array}{rcl}  r^2 - 3R^2
 &\hbox{for}& r \le R\\-\frac{2R^3}{r}& \hbox{for} & r \ge R\end{array} \right .
\end{eqnarray}
It is easily seen that the variation of this potential, or the
deviation of the Higgs field from the usual VEV, over the range
$r$ running from $R$ to $\infty$ is twice that over the range of
$r$ going from $0$ to $R$. Thus the potential variation from $r
\rightarrow \infty$ to $r=0$ is 3/2 times that from $r \rightarrow
\infty$ to $r=R$ (which we took as the average radius).

As we saw in Sec.~\ref{sec5}, for the case where gluons are
ignored, the Higgs field became zero at the average radius, $R$.
So, in this case, the central value of the Higgs field would be
\be \phi_{h}|_{r =0} = -\frac{1}{2}<\phi_h> = -\frac{1}{2}v \ee
i.e., opposite in sign and half the magnitude of the usual VEV
$v$. However, when we take into account the gluon part of the
binding (see Sec. 5.1), we only need the potential at the average
distance to be $4/9$ of what it was for the case of ignoring the
gluons. In this case we got the Higgs field at the average
distance $R = <r>$  to be $(1-4/9)v$ = $5v/9$. Then the field
strength at the center becomes \be \phi_{h}|_{r=0} = ( 1 - 3/2 *
4/9)v = v/3. \ee


For $r \le r_{inflection}$ the second derivative of the effective
potential $V_{eff}(\phi_h)$ for the Higgs field is negative, so that the effective
Higgs mass in this region is imaginary. We now want to get a typical
average value for this second derivative to be used in estimating the
effective imaginary Higgs mass in this range.

For orientation we note that, while the second derivative of
$V_{eff}(\phi_h)$ at the inflection point where $r=r_{inflection}$
is by definition just zero, we have that the Higgs field takes the
value $v/3$ at $r=0$ when gluons are included. Now, for the Higgs
field $\phi_h = v/3$, the second derivative of the $V_{eff}$ is
$-\frac{1}{3}$ times its value at the minimum of the effective
potential $V_{eff}$, i.e., where it is equal to the physical Higgs
mass squared. Thus the effective Higgs mass squared at the value
of the field in the central region of the bound state is
$-\frac{1}{3}m_h^2$, where $m_h$ is the physical Higgs mass.

 In order to get an estimate of the effective imaginary Higgs mass in the
region of $r$ going from $0$ to $r_{inflection}$, we may linearly
interpolate the second derivative as a function of $r$ but then
remember to weight the importance of the various $r$ regions with
the weight factor $r^2$. The first step in our crude estimate is
to approximate the second derivative as a linear function in the
distance from the center $r$,
  \be
  \frac{d^2V(r)}{d\phi_h^2} = -\frac{1}{3}m_h^2\left(1-\frac{r}{r_{inflection}}
\right). \label{d2V}
  \ee
Introducing the notation $ x=\frac{r}{r_{inflection}}$, we then
see that the average value of $ \frac{d^2V(r)}{d\phi_h^2}$,
weighted with $r^2$, in the range $r \in [0, r_{inflection}]$ is
\be <\frac{d^2V(r)}{d\phi_h^2}> = -\frac{1}{3} m_h^2
\frac{\int_0^1 x^2(1-x)dx}{\int_0^1 x^2dx} = -\frac{m_h^2}{12}.
\ee It is easily seen that indeed the effective mass squared
$m^2_{h \ eff}$ of the Higgs is just \be m^2_{h \ eff} = \frac{d^2
V(r)}{d \phi_h^2}. \ee So that for its average value we get \be
m_{h \ eff}^2 = - \frac{m_h^2}{12}. \ee

\section*{Appendix H, Bound state mass dependence on the number of
constituents}

In Sec.~\ref{intermediate} we need an estimate of the mass of the
10-constituent bound state rather than that of the 12-constituent
bound state, which we are requiring to be massless. We shall
therefore present here a first estimate of the form of the
dependence of the mass squared of our family of bound states on
the number of ($t$ and $\overline{t}$) constituents $\hat{Z} = Z +
1$.

We argued in Appendix B that the mass squared $M^2 =m^2_{bound}$
of the bound state should be an analytic function of the
``parameters'', such as $g_t$ or even, as we shall use here, of
$\hat{Z}$. In other words we shall assume that the mass squared of
the bound state $M^2(\hat{Z})$ is an analytic function of the
number of constituents $\hat{Z}$.

In the weak coupling approximation (i.e., $g_t$ and $\alpha_s$
small), the mass of the bound state becomes $M(\hat{Z}) \approx
m_t \hat{Z}$, since it is essentially given by adding the masses
of the constituents. This is a reasonable approximation for small
$\hat{Z}$ and thus we obtain \be M^2(\hat{Z}) \approx m_t^2
\hat{Z}^2 \ee as a valid approximation for small $\hat{Z}$.

Now, however, there is a binding energy term, which becomes bigger
and bigger as $\hat{Z}$ increases. The total potential energy of
the constituents is proportional to the number of interacting
pairs and is thus proportional to $\hat{Z}^2$ or strictly speaking
$\hat{Z}(\hat{Z}-1)$. Hence each constituent feels a potential
proportional to $\hat{Z}^2/\hat{Z} = \hat{Z}$ or strictly
$\hat{Z}(\hat{Z} - 1)/\hat{Z} = \hat{Z} -1$. At the same time the
average distance of the constituent from the center of the bound
state is diminished, as in the hydrogen atom, in the same
proportion. It follows that the binding energy per particle
becomes proportional to the square of this factor. So the total
binding energy of the bound state is proportional to
$\hat{Z}\hat{Z}^2$ or strictly $\hat{Z}(\hat{Z} - 1)^2$.

Thus we are led to the following Taylor expansion of
$M^2(\hat{Z})$: \be M^2(\hat{Z}) = (m_t\hat{Z} - A\hat{Z}^3
+\cdots)^2 = m_t^2 \hat{Z}^2(1 - B\hat{Z}^2 + \cdots) \ee or
strictly speaking \be M^2(\hat{Z}) = (m_t\hat{Z} -
A'\hat{Z}(\hat{Z} - 1)^2 +\cdots)^2 = m_t^2 \hat{Z}^2(1 -
B'(\hat{Z}-1)^2 + \cdots) \ee

The main point of the present article is to investigate the
hypothesis that the top-quark Yukawa coupling is fine-tuned, so as
to make the mass of the bound state with $\hat{Z} = 12$
constituents just zero. Imposing this requirement onto the above
Taylor expansion leads to a smooth ansatz of the form \be
M^2(\hat{Z}) =
m_t^2\hat{Z}^2\left(1-\left(\frac{\hat{Z}}{12}\right)^2\right)
\label{taylor} \ee or strictly speaking we should have \be
M^2(\hat{Z}) =
m_t^2\hat{Z}^2\left(1-\left(\frac{(\hat{Z}-1)}{11}\right)^2\right).
\ee

\begin{figure}[h]
\centering
\includegraphics[width=70mm]{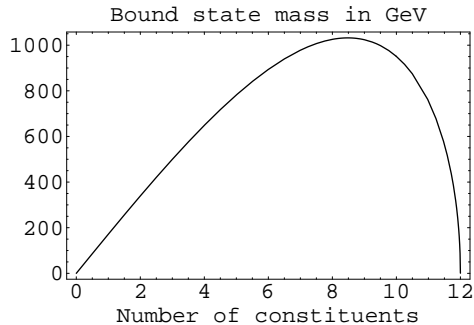}
\caption{Mass spectrum of bound states.} \label{2-f1}
\end{figure}

We now use the Taylor expansion (\ref{taylor}) to give a first
order estimate of the masses for the 11- and 10-constituent bound
states: \be m_{11} = \sqrt{11^2m_t^2\left(1-
\left(\frac{11}{12}\right)^2\right)} = 4.4m_t = 760 \ \mbox{GeV}
\ee while \be m_{10} = \sqrt{10^2m_t^2\left(1-
\left(\frac{10}{12}\right)^2\right)} = 5.5m_t = 950 \ \mbox{GeV}
\ee The full spectrum is shown in Fig.~\ref{2-f1}.

\section*{Appendix I: Gaussian wave function ansatz}

We shall now construct an ansatz for an approximation to the
multiparticle wave function for our system, consisting of the 6
top and 6 antitop particles, based on Gaussian functions. The main
purpose of this exercise is to confirm from a concrete model
ansatz the major part, namely, a factor 2 in the binding energy,
of the many body correction of Sec.~\ref{s9}.

The ansatz wave function for the $N$-particle system proposed here
is simply of the form \be \psi(\vec{x}_1,\cdots , \vec{x}_N) =
{\cal N}\prod_{i=1}^{N} \exp(-a_i \vec{x}_i^2). \label{ansatz} \ee
Of course, in our case of 12 constituents, we have $N=12$. The
idea then is to use the Hamiltonian based on the application of
the potential $V_{total}$ from Eq.~(\ref{Vtotal}) and the kinetic
energy summed over the $N$ particles, or we may simply use
$H=\sum_i H_i$ with $H_i$ taken from Eq.~(\ref{Hi}): \be H =
\sum_{i=1}^N \frac{\vec{p}_i^2}{2m_i} + \frac{1}{2}\sum_{i,j, \
i\neq j}V_{ij}. \ee Here $V_{ij} = \frac{A}{4\pi r_{ij}}$ is given
by Eq.~(\ref{Vijtotal}), with $r_{ij} = |\vec{x}_i - \vec{x}_j|$
being the distance between particle number $i$ and particle number
$j$.

The idea now is that we imagine to find the best possible wave
function of this form for the bound state system, by evaluating
the average energy in such an ansatz state as a function of the
parameters $a_i$. In practice, for our symmetric case, we obtain
the same value for all the $N$ $a_i$'s and simply minimize the
energy with respect to their common value $a_i = a$. Using our
Gaussian ansatz, we obtain \be <H_i> = \frac{3a}{2m_t} -
(N-1)\frac{A}{4\pi}\sqrt{\frac{a}{\pi}} \ee for the expectation
value of the single particle Hamiltonian $H_i$. Minimizing this
energy determines our variational parameter to be \be a = (N-1)^2
\left(\frac{A}{4\pi}\right)^2 \frac{m_t^2}{9\pi}. \ee This gives,
using  \be <H> = N <H_i> = -\frac{N(N-1)^2A^2m_t}{96\pi^3} =
-\frac{(e^2_{t \overline{t}} + 4 g_t^2)^2}{32\pi^2}
\frac{4}{3\pi}m_t \label{Egaussian} \ee for the factorizable
Gaussian wave function estimate of (minus) the binding energy of
the bound state, where we have substituted $N=12$ and the
expression for $A$ from Eq.~(\ref{Vijtotal}).

We can now compare this value (\ref{Egaussian}) for the binding
energy\footnote{We note that this value is in agreement with the
calculation of the many body effect in a recent paper
\cite{Shuryak} by Kuchiev, Flambaum and Shuryak, when the
correction by a factor of 2 mentioned in footnote 1 and the
reduced mass factor of 11/12 are taken into account. In fact it
means that in the notation of Ref.~\cite{Shuryak} we would obtain
$k=1/6\pi\approx 0.053$, while in their variational calaculation
they obtain $ k=25/512 \approx 0.049$.} with the ``Bohr model''
approximation $E_{binding}$ of Eq.~(\ref{Ebinding}): \be - <H> =
E_{binding}*\frac{1}{2}*\frac{12}{11}*\frac{8}{3\pi} =
\frac{E_{binding}}{2.16}. \label{216factor} \ee Thus we have
basically reproduced the expected main reduction in the binding
energy by a factor of 2 due to many body effects. We note that the
extra factor of $\frac{11}{12}$ arises from the reduced mass of a
single quark moving relative to the other 11 quarks after removing
the center of mass motion. The final factor of $\frac{8}{3\pi}$
corresponds to the reduction in the Bohr model binding energy
obtained by using a Gaussian form rather than the exact Bohr wave
function.

\section*{Appendix J: Phase transition in bound state calculation}

We shall illustrate the possibility for the appearance of a phase
transition in the bound state calculation, which can explain the
disagreement between the present paper and Ref.~\cite{Shuryak}.
For reasons of tractability we do not consider the genuine bound
state calculation, but rather a toy model that simulates a
continuous material made from such bound states and extended to
infinity.

Really our toy model is a material with an {\em a priori} fixed
density of both top and antitop quarks. But then the idea is to
adjust the density of top and antitop quarks so as to correspond
to the situation in which the bound states just fill the space
completely without overlapping.

It is important that  we treat top and antitop quarks as different
species of the same type of particle, which are separately
conserved. So it only matters how many top or antitop quarks there
are together in states with a given momentum. The number of
possible states for a given momentum is denoted by $N_{sp} = 2 * 2
* N_c$. Here $N_c$ is the number of colors. So $N_{sp} =12$ is
the case of interest for nature and is the value we use below. As
part of our toy model we ignore annihilation completely, so that
particles and antiparticles are separately conserved. Then we take
the Fermi momentum $p_f$ as an ansatz parameter. From this alone
we can derive the density of the top quarks and antitop quarks in
the ansatz state: \be \rho = N_{sp}\frac{4\pi p_f^3}{3(2\pi)^3}.
\ee Their energy density is then \be ``\hbox{energy density of
fermions}'' =\frac{N_{sp}}{(2\pi)^3} \int_0^{p_f}4\pi p^2
\sqrt{p^2 +m^2} dp. \ee

Now the fermion mass comes from the Higgs field and we have \be m
= g_t <\phi_h>/\sqrt{2}. \ee The potential energy for the Higgs
field $\phi_h$ is of the form \be V_{eff}(\phi_h) = -\frac{1}{2}
|m_{hb}|^2 \phi_h^2 + \frac{\lambda}{8} \phi_h^4. \ee For use in
the present Appendix, we introduce the effective potential
normalized to be just zero at the (usual) minimum:
\begin{eqnarray}
V_{effnorm}(\phi_h) &= &-\frac{1}{2} |m_{hb}|^2 \phi_h^2 +
\frac{\lambda}{8}\phi_h^4 - V_{eff}(<\phi_h>) \\
& = & -\frac{1}{2} |m_{hb}|^2 \phi_h^2
+ \frac{\lambda}{8}\phi_h^4 + \frac{|m_{hb}|^4}{2\lambda}.
\end{eqnarray}
We consider the approximation in which the Higgs field $\phi_h$ is
taken to have a constant value inside the bound state. So the
kinetic energy of the Higgs field can be ignored and
thus the total energy density $U$ in our toy model ansatz
becomes
\begin{eqnarray}
U &=& \frac{N_{sp}}{2\pi^2}
\int_0^{p_f} p^2 \sqrt{p^2 + (g_t \phi_h/\sqrt{2})^2}dp + V_{effnorm}(\phi_h)\\
&=& \frac{N_{sp}}{32\pi^2}\left(\frac{(g_t\phi_h)^4}{4}
\left[\log \frac{(g_t\phi_h)^2}{2}-2\log(p_f+\sqrt{p_f^2+(g_t\phi_h)^2/2})\right]
\right)\nonumber \\
& & +\frac{N_{sp}}{32\pi^2}\left(2p_f(2p_f^2+\frac{(g_t\phi_h)^2}{2})
\sqrt{p_f^2+(g_t\phi_h)^2/2}\right) +  V_{effnorm}(\phi_h).
\end{eqnarray}

The Fermi momentum $p_f$ really determines the density of quarks
or antiquarks and thus---if bound states are effectively
present---also the density of the bound states. Now we want to
adjust the density in such a way as to crudely represent the fact
that the space is filled up with bound states, so that in every
point of space there is just one of the bound states present. That
is to say we must adjust the Fermi-momentum $p_f$ to such a value
that we achieve this density corresponding to totally filling
space with bound states. After adjusting $p_f$ in such a way, we
can obtain the mass or rather the energy of the bound state by
using the fact that the number of bound states per unit volume is
\be \frac{4\pi}{3(2\pi)^3}p_f^3 =\frac{p_f^3}{6\pi^2}. \ee So the
mass or rather the energy of the potential bound state (ENBS) is
\be ENBS = \frac{6\pi^2}{p_f^3} U.\label{ENBS} \ee

Now the density needed to have filled space with the bound states
can---crudely at least---be found by minimizing the bound state
energy ENBS with respect to the variable determining the density,
i.e., with respect to $p_f$. The argument for this runs as
follows:

1) If we make an ansatz ``material'' with a lower density of bound
states than there is place for, then each bound state can be
imagined to be surrounded by a little piece of essentially vacuum
the energy of which must be added to the value ENBS as calculated
from (\ref{ENBS}). Now we have normalized the effective potential
$V_{effnorm}(\phi_h)$ by making it vanish at its minimum. So the
pieces of new vacuum in any ansatz will have positive energy.
Thus, if we make the density in the ansatz too low, the result for
ENBS will always be larger than the true bound state energy.

2) On the other hand if we make an ansatz with a too high density
so that the bound states get squeezed together, this will also
cause the energy per bound state ENBS to increase compared to that
of a free bound state.

So we see that the energy formally calculated from an ansatz ENBS
will be bigger both when the density is higher and when it is
lower than the one corresponding to the bound states just touching
or filling the space. This then means that there must be a minimum
in the energy per bound state ENBS as a function of the density
parameter $p_f$.

Since we are working in the approximation of letting the Higgs
field be constant inside the bound state, we really just want to
adjust this Higgs field $\phi_h$ so as to minimize the energy of
the bound state. Combined with the above-mentioned adjustment of
$p_f$, we end up with the rule that we shall adjust both
parameters $p_f$ and $\phi_h$ so as to minimize the expression
(\ref{ENBS}) for ENBS. Then we should obtain, in our ansatz
approximation, the right mass or rather energy for the bound state
if there is a bound state. If there is no binding, we should get
the energy of the $N_{sp}$ ``constituents'' that were meant to be
bound. In the case of a potential bound state made from $N_{sp} =
12$ top or antitop quarks, this constituent energy would of course
be $N_{sp}$ times the top-quark mass (or energy, but we expect
that the speed would be low in our ansatz).

The main point of this Appendix is that {\em the mass or energy of
the bound state appears as the result of taking a minimum so that
it will not normally be a nice analytical function of the
parameters that are input into the calculation such as $g_t$, but
rather tends to have a kink as a function of the inputs.}

Without taking into account on which side of the ``phase
transition'' a given value of $g_t$ may lie, one can {\em a
priori} make a severe error in the calculation. According to our
toy model, the correct side of the phase transition is determined
by the question of whether or not the Higgs field in the region of
the potential bound state has been pushed so much as to deviate
strongly from its value in the usual vacuum. The calculation in
Ref.~\cite{Shuryak} has been made on the small $g_t$ side of the
phase transition, where to a very good approximation we have the
usual vacuum with the usual 246 GeV Higgs field expectation value.
On the other hand, in this paper we have worked in the regime
where we take the Higgs field in the interior of the hypothesized
bound state to deviate significantly from that in the usual
vacuum. Indeed the typical field value inside the bound state in
our calculation is rather small. So we have worked on the large
$g_t$ side of the phase transition.

We now present the results of our toy model calculation, which
exhibit the existence of such a phase transition. Here we use a
Higgs mass of $m_h = 115$ GeV. The results obtained for the mass
or really the energy of the potential bound state ENBS are plotted
in Fig.~2 as a function of $g_t$.
\begin{figure}[h]
\centering
\includegraphics[width=70mm]{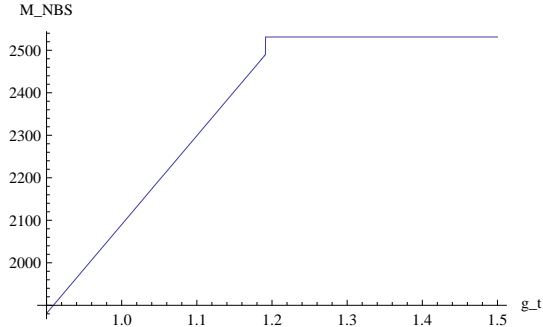}
\caption{Energy or mass of the ``bound state'' in GeV in our toy
model as a function of the top-quark Yukawa coupling $g_t$.}
\label{3-f1}
\end{figure}

They were calculated by simply minimizing, for each choice of the
Yukawa coupling $g_t$, the value of ENBS as given by (\ref{ENBS})
with respect to both variables, the Fermi momentum $p_f$ and the
Higgs field (in the interior of the bound state) $\phi_h$. The
little step in the figure is an artifact of the calculational
accuracy, but the kink is of course due to the minimum giving the
smallest ENBS jumping discontinuously at $g_t = 1.191$. Indeed the
minimum jumps from $(\phi_h, p_f)$ = (246, 0.18 GeV) to (0, 211
GeV).

This jumping is partly illustrated by Fig.~3, where the potential
bound state mass or rather energy ENBS is plotted as a function of
$\phi_h$, when the latter is imposed as the approximate value of
the Higgs field inside the bound state region. It means that for
every $\phi_h$ value the function ENBS from (\ref{ENBS}) has been
minimized with respect to effectively the density of bound states,
meaning minimization with respect to $p_f$. Figure 3 is made for
the specific value $g_t = 1.191$, which is the phase transition
value. This is reflected by the fact that you see two essentially
degenerate minima in Fig.~3.
\begin{figure}[h]
\centering
\includegraphics[width=70mm]{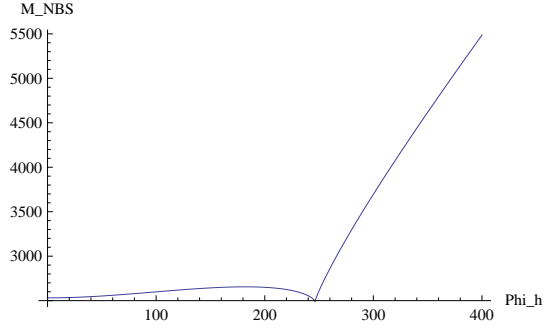}
\caption{Energy or mass of the bound state in GeV in our toy model
as a function of the imposed $\phi_h$ value, but with $p_f$
adjusted by minimization. The top-quark Yukawa coupling is chosen
to have the phase transition value $g_t = 1.191$.} \label{4-f1}
\end{figure}

For $g_t$ greater than the phase transition value of 1.191, the
mass or energy remains constant as the Yukawa coupling $g_t$
increases. This means that the binding gets stronger and stronger,
in as far as the binding energy is really \be ``\hbox{binding}''=
\frac{N_{sp} g_t}{\sqrt{2}}  246 \ \mbox{GeV} - ENBS. \ee Thus,
for example, in our toy model the binding energy becomes equal to
half the mass of the constituents for $g_t = 2.42$. According to
our discussion in Sec.~\ref{justification}, this is the formal
requirement for a massless bound state. Thus, in the bad
approximation of ignoring the exchange of eaten Higgses, gluon
exchange etc.~and even {\em taking the Higgs field inside the
bound state as constant}, we obtain $g_t = 2.42$ as the value of
the Yukawa coupling which gives a massless bound state in our toy
model.

For the case of $g_t$ less than the phase transition value of 1.191,
we get a very small value for $p_f$ compared to our own results from
the Bohr atom approximation. We get $p_f \sim 0.18$ GeV rather than
of order $g_t^2 m_t$. This very small value of $p_f$ may be interpreted as
supporting (as does Fig.~2 for $g_t < 1.191$) the result of Kuchiev
et al.~according to which the system does not bind. Completely zero binding
would correspond to each particle standing still and well separated from
each other, which would imply a very low density and thus correspond
to $p_f = 0$.

\end{document}